\documentclass[twocolumn,showpacs,aps,prd,superscriptaddress,nofootinbib]{revtex4}

\usepackage[pdftex]{graphicx}
\usepackage{dcolumn}
\usepackage{amsmath}
\usepackage{epsfig}

\def\lsim{\mathrel{\rlap{\lower4pt\hbox{\hskip1pt$\sim$}}
    \raise1pt\hbox{$<$}}}                
\def\gsim{\mathrel{\rlap{\lower4pt\hbox{\hskip1pt$\sim$}}
    \raise1pt\hbox{$>$}}}                

\begin{document}

\title{
  {\large 
    \bf \boldmath 
    Prospects for the Measurement of the Unitarity Triangle Angle $\gamma$
    from $B^0 \to DK^+\pi^-$ Decays
  }
}

\author{Tim Gershon}
\affiliation{Department of Physics, University of Warwick, Coventry CV4 7AL, United Kingdom}
\author{Mike Williams}
\affiliation{Physics Department, Imperial College London, London, SW7 2AZ, United Kingdom}

\date{\today}

\begin{abstract}
The potential for a precise measurement of the Unitarity Triangle angle 
$\gamma$ in future experiments from the decay $B^0 \rightarrow DK^{*0}$ is 
well-known. 
It has recently been suggested that the sensitivity can be significantly
enhanced by analysing the $B^0 \rightarrow DK^+\pi^-$ Dalitz plot to extract 
amplitudes relative to those of the flavour-specific decay 
$B^0 \rightarrow D_2^{*-}K^+$. 
An extension to this method which includes the case where the neutral $D$ meson
is reconstructed in suppressed final states is presented.  
The sensitivity to $\gamma$ is estimated using this method and compared to 
that obtained using the $B^0 \rightarrow DK^{*0}$ decay alone.  
Experimental effects, such as background contamination, are also considered.
This approach appears to be a highly attractive addition to the
family of methods that can be used to determine $\gamma$.
\end{abstract}


\maketitle

\section{Introduction}
\label{sec:intro}

The quark flavour sector of the Standard Model of particle physics,
described by the Cabibbo-Kobayashi-Maskawa (CKM)
quark mixing matrix~\cite{Cabibbo:1963yz,Kobayashi:1973fv},
gives a successful description of all current experimental measurements of
quark flavour-changing interactions.  
It also provides an excellent laboratory to search for effects of physics
beyond the Standard Model (see, for example, Refs.~\cite{Buchalla:2008jp,Browder:2008em,Antonelli:2009ws}).
A critical element of this programme is the precise measurement of 
the angle ${\gamma = {\rm arg}\left(-V_{ud} V^*_{ub}/V_{cd}V^*_{cb}\right)}$ of
the Unitarity Triangle formed from elements of the CKM matrix.

A method to measure $\gamma$ with negligible theoretical uncertainty was
proposed by Gronau, London and Wyler
(GLW)~\cite{Gronau:1990ra,Gronau:1991dp}.
The original method uses $B \to DK$ decays, 
with the neutral $D$ meson reconstructed in $CP$ eigenstates.
The method can be extended to use $D$ meson decays to 
any final state that is accessible to both $D^0$ and $\overline{D}{}^0$,
in particular doubly-Cabibbo-suppressed decays such as
$K^+\pi^-$~\cite{Atwood:1996ci,Atwood:2000ck} have been noted to provide
enhanced sensitivity to $CP$-violation effects.

The use of neutral $B$ decays is particularly interesting since the two
contributing amplitudes are more similar in magnitude, so that direct 
$CP$-violation effects may be enhanced relative to those in charged $B$ decays.
The decay ${B^0 \to DK^{*0}}$ is especially advantageous since the 
charge of the kaon in the ${K^{*0} \to K^+\pi^-}$ decay
unambiguously tags the flavour of the decaying $B$ meson, 
obviating the need for time-dependent analysis~\cite{Dunietz:1991yd}.
This appears to be one of the most promising channels for LHCb 
to make a precise measurement of $\gamma$~\cite{Akiba:2007zz,Akiba:2008zz,Nardulli:2008}.

The approach that has mainly been considered until recently is a
quasi-two-body analysis of ${B^0 \to DK^{*0}}$.
In this analysis, the contributions from other resonances in the 
${B^0 \to DK^+\pi^-}$ 
Dalitz plot that interfere with the $K^{*0}$ within the selected
mass window  are handled by the introduction of an additional hadronic
parameter~\cite{Gronau:2002mu}.
This parameter, normally denoted by $\kappa$, takes values between 0 and 1
where 0 implies that all sensitivity to $\gamma$ is lost, and the limit of 1
is reached in the case that no amplitudes other than $DK^{*0}$ contribute.
Estimates suggest that ${0.9 < \kappa < 1.0}$ for a $K^{*0}$ mass window of 
$\pm 50 \ {\rm MeV}$~\cite{Pruvot:2007yd}.

Recently it has been noted that the natural width of the $K^*$ meson
can be used to enhance the sensitivity to the $CP$-violating phase $\gamma$ 
through analysis of the ${B^0 \to DK^+\pi^-}$ Dalitz
plots~\cite{Bediaga,Gershon:2008pe}.
By comparison of the Dalitz-plot distributions of events in the cases where 
the neutral $D$ meson is reconstructed in flavour-specific and $CP$-eigenstate
modes, the complex amplitudes of the $DK^{*0}$ decays can each be determined 
relative to the flavour-specific $D_2^{*-}K^+$ amplitude.  This allows for 
a direct extraction of $\gamma$ from the difference in amplitudes,
rather than from the rates.

In this paper we extend the method proposed in Ref.~\cite{Gershon:2008pe} to
include also the case where the neutral $D$ meson is reconstructed in
suppressed final states.
This allows us to make a direct comparison of the sensitivity to $\gamma$ 
between the quasi-two-body analysis and the Dalitz-plot analysis.
We also study possible systematic effects that may limit the sensitivity of
the analysis, including uncertainties on the correct composition of the 
Dalitz-plot model and a brief discussion of experimental effects.

The remainder of the paper is organised as follows:
in Section~\ref{sec:method} we give an overview of the method;
in Section~\ref{sec:model} we describe the Dalitz-plot model used for our
study; 
in Sections~\ref{sec:q2b-results} and~\ref{sec:dalitz-results} we present the
results we obtain in the quasi-two-body analysis and in the Dalitz-plot
analysis, respectively;
in Section~\ref{sec:effects} we discuss how experimental effects can be
handled,
before summarising in Section~\ref{sec:summary}.

\section{Method}
\label{sec:method}

\begin{figure}[!htb]
  \includegraphics[height=0.49\columnwidth,angle=90]{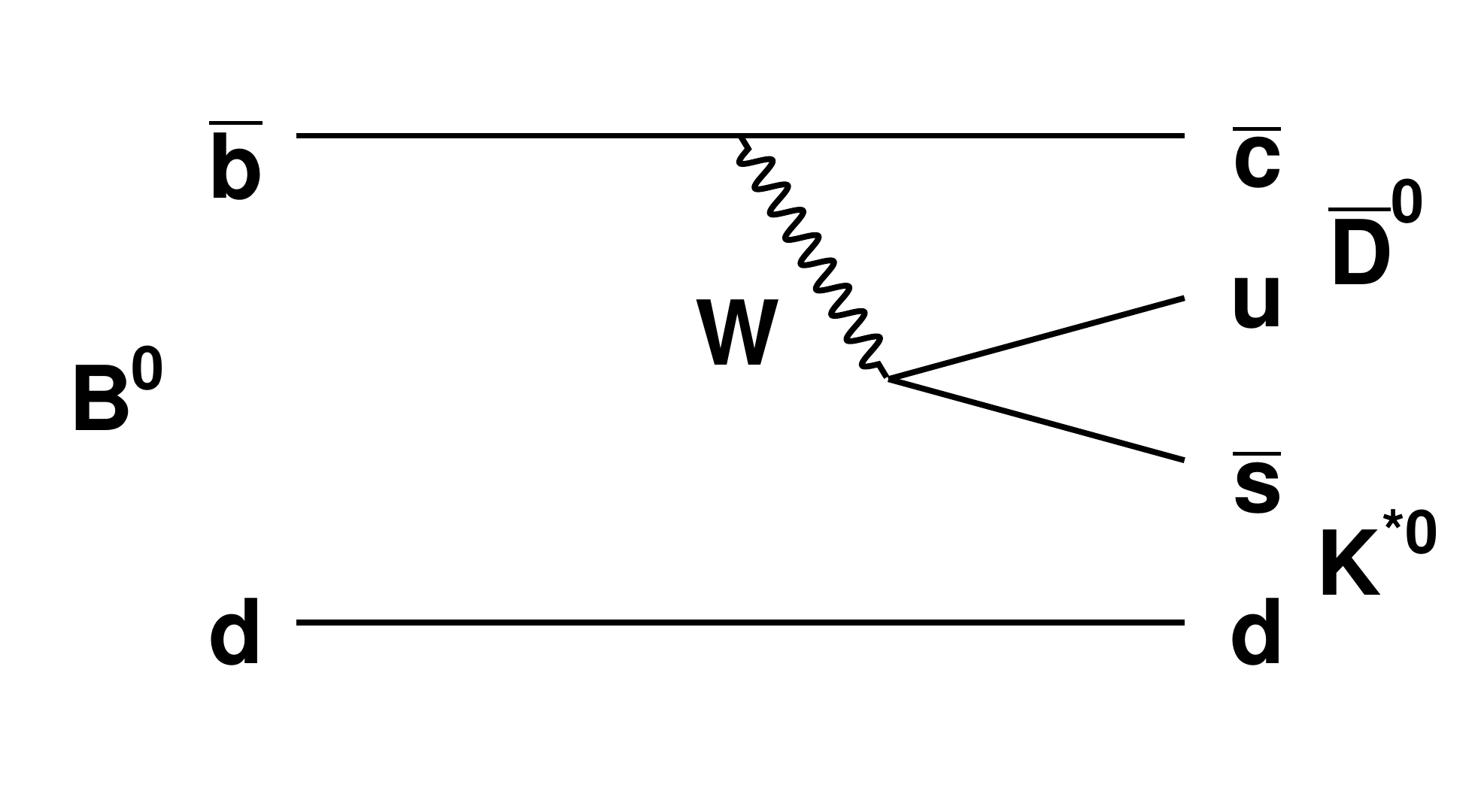}
  \includegraphics[height=0.49\columnwidth,angle=90]{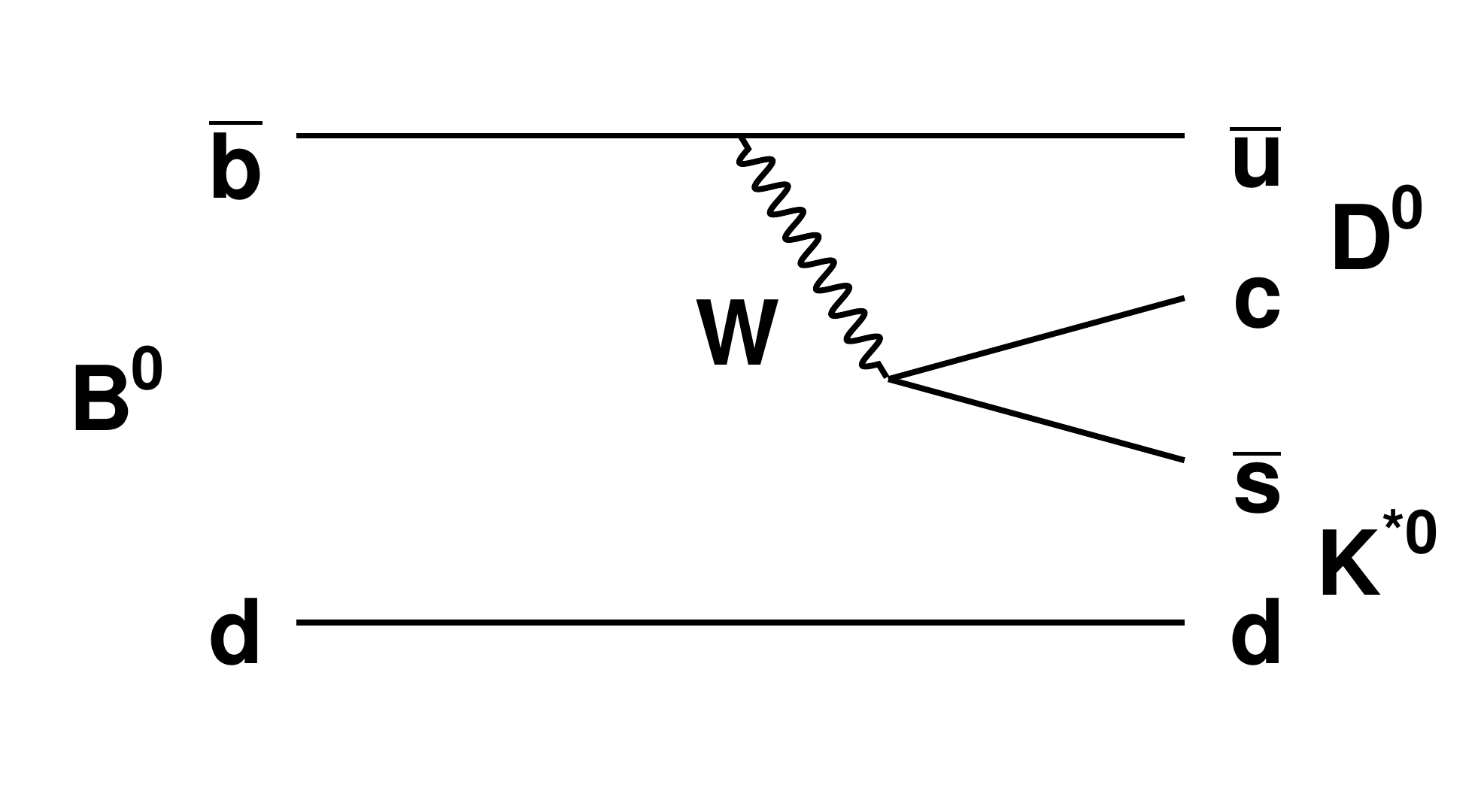}
  \caption{
    Feynman diagrams for $B^0 \to DK^{*0}$, via 
    (left) a $\bar{b} \to \bar{c} u \bar{s}$ transition
    and (right) a $\bar{b} \to \bar{u} c \bar{s}$ transition.    
  }
  \label{fig:feynman}
\end{figure}

In this Section we describe the various methods that can be used to extract
$\gamma$ from $B \to DK^{(*)}$.
We first review the quasi-two-body approach, 
and then recap the recently proposed Dalitz-plot technique~\cite{Gershon:2008pe},
before describing the extension to the method used in our study.
All methods exploit the interference between the two tree-level amplitudes
shown in Fig.~\ref{fig:feynman}.
Conventionally, the ratio of magnitudes of these two amplitudes is referred to
as $r_B$, while their strong phase difference is labelled $\delta_B$.

The GLW method~\cite{Gronau:1990ra,Gronau:1991dp} of extracting $\gamma$ uses
the following rates and asymmetries in $B \to DK^{(*)}$ decays:
\begin{eqnarray}
  R_{\pm} & = & 
  \frac{
    \Gamma(\overline{B} \to D_{\pm}\overline{K}{}^{(*)}) + \Gamma(B \to D_{\pm}K^{(*)})
  }{
    \Gamma(\overline{B} \to D_{\rm fav}\overline{K}{}^{(*)}) + \Gamma(B \to D_{\rm fav}K^{(*)})
  } \nonumber \\
  & = &
  1 + r_B^2 \pm 2 \, r_B \cos(\delta_B) \cos(\gamma) \, ,\\
  A_{\pm} & = & 
  \frac{
    \Gamma(\overline{B} \to D_{\pm}\overline{K}{}^{(*)}) - \Gamma(B \to D_{\pm}K^{(*)})
  }{
    \Gamma(\overline{B} \to D_{\pm}\overline{K}{}^{(*)}) + \Gamma(B \to D_{\pm}K^{(*)})
  } \nonumber \\
  & = &
  \frac{\pm 2 \, r_B \sin(\delta_B) \sin(\gamma)}{R_{\pm}} \, .
\end{eqnarray}
Here, $\overline{B}$ ($B$) is used to refer to 
either $B^-$ or $\overline{B}{}^0$ ($B^+$ or $B^0$), 
while $D_{\pm}$ refers to a neutral $D$ meson reconstructed in a $CP$-even
({\it eg.} $K^+K^-$) or $CP$-odd ({\rm eg.} $K_S^0\pi^0$) final state, and
$D_{\rm fav}$ refers to a neutral $D$ meson reconstructed in a favoured,
quasi-flavour-specific ({\it eg.} $D^0 \to K^-\pi^+$) final state.
Note that experimentally it is convenient to measure $R_{\pm}$ normalised to
an equivalent double ratio from $B \to D\pi$ or $B \to D\rho$ decays.

Since $R_{+}A_{+} + R_{-}A_{-} = 0$, the above four observables give three
independent constraints on the three parameters $\gamma$, $r_B$ and $\delta_B$. 
This is sufficient to solve the system up to an eightfold ambiguity.
However, when measurements are performed in a hadronic environment as at LHCb,
the reconstruction of the $CP$-odd final states becomes a significant
experimental problem.  Therefore, additional observables are required.

A solution is to include doubly-Cabibbo-suppressed $D$ meson decays,
as first suggested by Atwood, Dunietz and Soni
(ADS)~\cite{Atwood:1996ci,Atwood:2000ck}. 
Due to the fact that the $D$ meson decays to $K^\pm\pi^\mp$ are not truly
flavour-specific, but include a suppressed contribution which is given by
$r_De^{i\delta_D}$ relative to the favoured decay amplitude,\footnote{
  Note that the sign convention for $\delta_D$ used in this paper is opposite
  to that used in most of the literature on $\gamma$ measurements.
}
enhanced $CP$-violation effects can occur.
The rates and asymmetries of the $B\to DK^{(*)}$ decays to the suppressed
final states are then given by
\begin{eqnarray}
  R_{\rm ADS} & = & 
  \frac{
    \Gamma(\overline{B} \to D_{\rm sup}\overline{K}{}^{(*)}) + \Gamma(B \to D_{\rm sup}K^{(*)})
  }{
    \Gamma(\overline{B} \to D_{\rm fav}\overline{K}{}^{(*)}) + \Gamma(B \to D_{\rm fav}K^{(*)})
  } \nonumber \\
  & = & r_B^2 + r_D^2 + 2\,r_B\,r_D\cos(\delta_B-\delta_D)\cos(\gamma) \, ,\\
  A_{\rm ADS} & = & 
  \frac{
    \Gamma(\overline{B} \to D_{\rm sup}\overline{K}{}^{(*)}) - \Gamma(B \to D_{\rm sup}K^{(*)})
  }{
    \Gamma(\overline{B} \to D_{\rm sup}\overline{K}{}^{(*)}) + \Gamma(B \to D_{\rm sup}K^{(*)})
  } \nonumber \\
  & = & \frac{2\,r_B\,r_D\sin(\delta_B-\delta_D)\sin(\gamma)}{R_{\rm ADS}} \, .
\end{eqnarray}
Since the hadronic parameters of the $D$ decay ($r_D$ and $\delta_D$) can be
determined independently~\cite{Rosner:2008fq,Asner:2008ft},
these measurements provide two additional linearly independent constraints
that can be used in combination with the GLW observables to obtain bounds on
the three unknown parameters $\gamma$, $r_B$ and $\delta_B$.
With the decay modes $D \to K^\pm\pi^\mp$,
the ADS observables are well suited to reconstruction in a hadronic
environment. Consequently, one of the most promising strategies for the
tree-level determination of $\gamma$ at LHCb is that from the combination of
measurements of $R_{+}$, $A_{+}$, $R_{\rm ADS}$ and $A_{\rm ADS}$ in 
charged ${B \to DK}$ decays~\cite{LHCb-2009-011} 
or neutral ${B\to DK^*}$ decays~\cite{Akiba:2007zz,Nardulli:2008}.

The finite width of the $K^{*0}(892)$ resonance leads to additional 
complications in the analysis of $B\to DK^*$ decays, since other contributions
to the $B \to DK\pi$ Dalitz plot can affect the population within the $K^*$
mass window.  This can be handled by making the following
substitutions~\cite{Gronau:2002mu}:
\begin{eqnarray}
  r_B \longrightarrow r_S & = & 
  \sqrt{\frac{\int_{\rm DP} \left| A_u \right|^2 d\vec{x}}{\int_{\rm DP} \left| A_c \right|^2 d\vec{x}}} \, ,\\
  e^{i\delta_B} \longrightarrow \kappa e^{i\delta_S} & = & 
  \frac{
    \int_{\rm DP} \left| A_u \right|  \left| A_c \right| e^{i\delta}d\vec{x}
  }{
    \sqrt{\int_{\rm DP} \left| A_u \right|^2 d\vec{x}\int_{\rm DP} \left| A_c \right|^2 d\vec{x}}
  } \, ,
\end{eqnarray}
where $A_c$ and $A_u$ are respectively the amplitudes carrying the phase of the
${\bar{b} \to \bar{c} u \bar{s}}$ (Fig.~\ref{fig:feynman}(left)) and of the 
${\bar{b} \to \bar{u} c \bar{s}}$ (Fig.~\ref{fig:feynman}(right)) transitions, 
and $\delta$ is the strong phase difference between them, all as functions of
the Dalitz-plot position $\vec{x}$.  The integrals are over the region of the
Dalitz plot that is defined as the $K^*$ mass window.
In the limit that $DK^*$ is the only contribution in this window, 
$r_S \longrightarrow r_B$, $\delta_S \longrightarrow \delta_B$ and 
$\kappa \longrightarrow 1$.

With this treatment, the two hadronic parameters associated with the $DK^*$
decay ($r_B$ and $\delta_B$) are replaced with two effective parameters ($r_S$
and $\delta_S$) and a new unknown ($\kappa$) is introduced.
Since the combination of ($CP$-even) GLW and ADS observables provides four
linearly independent measurements, it is possible to determine $\kappa$ from
the data together with $\gamma$, $r_B$ and $\delta_B$.  
Alternatively, external input, either theoretical or experimental, could be
used to constrain $\kappa$~\cite{Pruvot:2007yd}.

We refer to the extraction of $\gamma$ using the approach outlined above as
the quasi-two-body analysis. 
The addition of ADS observables helps to resolve two of the ambiguities of the
GLW approach~\cite{Soffer:1999dz};
however, the effectiveness of this depends on the values of $r_D$, $\delta_D$
and $\delta_B$, as well as the statistical sensitivity.
Moreover, the sensitivity to $\gamma$ depends on the values of the unknown
hadronic parameters, particularly $\delta_B$.

\begin{figure*}[!htb]
  \includegraphics[height=0.82\columnwidth,angle=90]{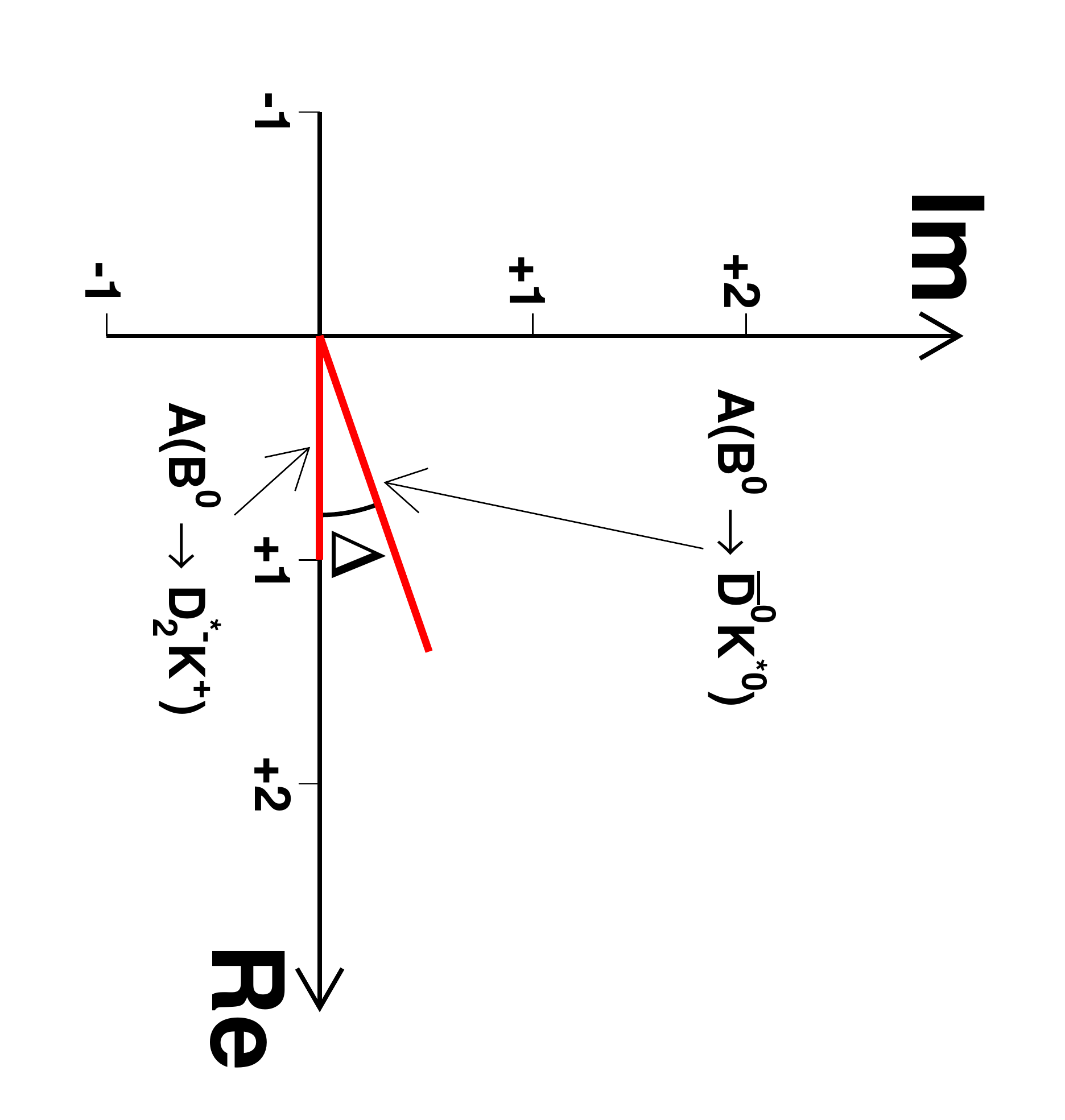}
  \hspace{8mm}
  \includegraphics[height=0.82\columnwidth,angle=90]{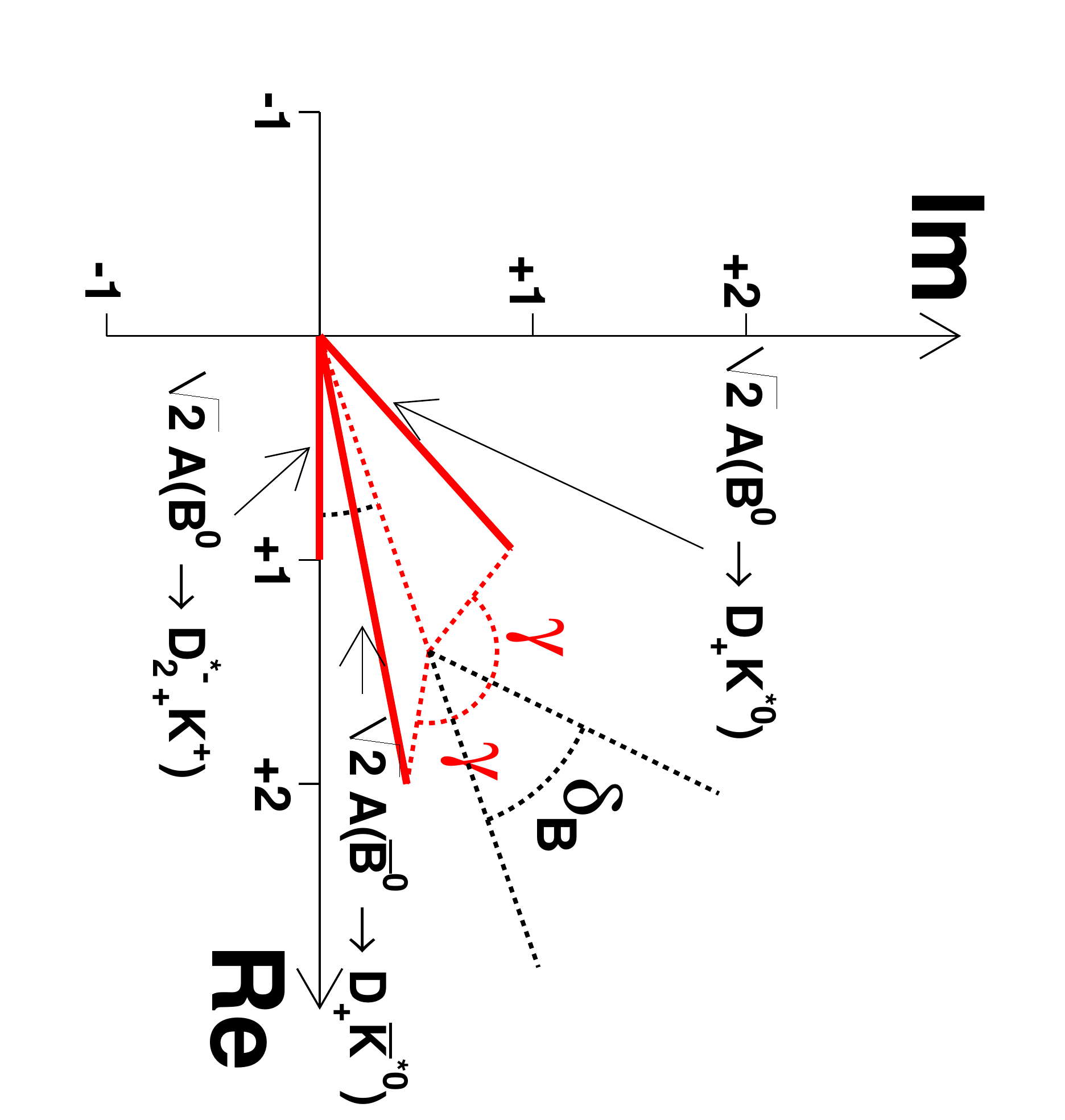}
  \caption{
    Argand diagrams illustrating the measurements of relative amplitudes 
    and phases from analysis of the Dalitz plots of 
    (left) $\overline{D}{}^0K^+\pi^-$ and (right) $D_{+}K^+\pi^-$.
    In these illustrative examples the following values are used:
    $\varrho = 1.5$, $\Delta = 20^\circ$, $\gamma = 75^\circ$,
    $\delta_B = 45^\circ$ and $r_B = 0.4$.
    \label{fig:argand}
  }
\end{figure*}

The recently proposed $B \to DK\pi$ Dalitz-plot analysis~\cite{Gershon:2008pe}
exploits the presence of the $B \to D_2^{*}K$ contribution that serves as a
reference amplitude, since the flavour of the neutral $D$ meson produced in 
$D_2^{*\pm} \to D\pi^\pm$ is tagged by the charge of the accompanying pion.
Considering flavour-specific $D$ mesons, we can define the $B\to DK^*$
amplitude relative to this reference, as illustrated in
Fig.~\ref{fig:argand}(left),
\begin{equation}  
  \label{eq:flavSpecAmpRatio}
  \frac{A(B^0 \to \overline{D}{}^0K^{*0})}{A(B^0 \to D_2^{*-}K^+)} = 
  \varrho e^{i\Delta} \, .
\end{equation}
Note that in Eq.~\ref{eq:flavSpecAmpRatio} and throughout the discussion below
we neglect factors of ${A(K^{*0} \to K^+\pi^-)}$ and 
${A(D_2^{*-} \to \overline{D}{}^0\pi^-)}$ that formally should appear in the
numerator and denominator respectively, since they eventually cancel in the
observables of interest.

Considering now $CP$-even $D$ mesons, using the convention 
$D_\pm = \frac{1}{\sqrt{2}}\left( D^0 \pm \overline{D}{}^0 \right)$,
we find (see Fig.~\ref{fig:argand}(right))
\begin{equation}  
  \label{eq:cpAmpRatio}
  \frac{\sqrt{2} A(B^0 \to D_+K^{*0})}{\sqrt{2} A(B^0 \to D_{2\,+}^{*-}K^+)} = 
  \varrho e^{i\Delta} \left( 1 + r_B e^{i(\delta_B+\gamma)} \right) \, ,
\end{equation}
where $D_{2\,+}^{*-}$ denotes that the neutral $D$ meson produced in the
decay of the $D_2^{*-}$ is reconstructed in a $CP$-even eigenstate.
Thus, we find~\cite{Gershon:2008pe}
\begin{eqnarray}
  x_+ + i y_+ & = & r_B e^{i(\delta_B + \gamma)} \nonumber \\
  & = & \frac{
    (\sqrt{2}A(D_{+}K^{*0}))/(\sqrt{2}A(D_{2\,+}^{*-}K^+))
  }{
    (A(\overline{D}{}^0K^{*0}))/(A(D_2^{*-}K^+))
  } - 1 \nonumber \\
  & = & \frac{\sqrt{2}A(D_{+}K^{*0})}{A(\overline{D}{}^0K^{*0})} - 1 \, ,
  \label{eq1}
\end{eqnarray}
where the variables $\left( x_+, y_+ \right)$ are the same as those used in
the analysis of $B \to DK$ with $D \to K_S^0\pi^+\pi^-$
decays~\cite{Poluektov:2006ia,Aubert:2008bd}.
Constraints on $x_- + i y_- = r_B e^{i(\delta_B - \gamma)}$ are likewise
obtained from equivalent expressions for the charge-conjugate
$\overline{B}{}^0$ decays.
The extraction of $\gamma$ from this Dalitz-plot analysis with only a single
unresolvable ambiguity ($\gamma \longrightarrow \gamma+\pi, \delta_B
\longrightarrow \delta_B + \pi$) is possible using only $CP$-even $D$ decays.
Consequently, we restrict our discussion to $CP$-even $D$ decays, since those
are experimentally accessible in a hadronic environment; however, we note that 
$CP$-odd decays give similar expressions, but with the right-hand side of the
last two relations of Eq.~\ref{eq1} multiplied by a minus sign.

\begin{figure*}[!htb]
  \includegraphics[height=0.82\columnwidth,angle=90]{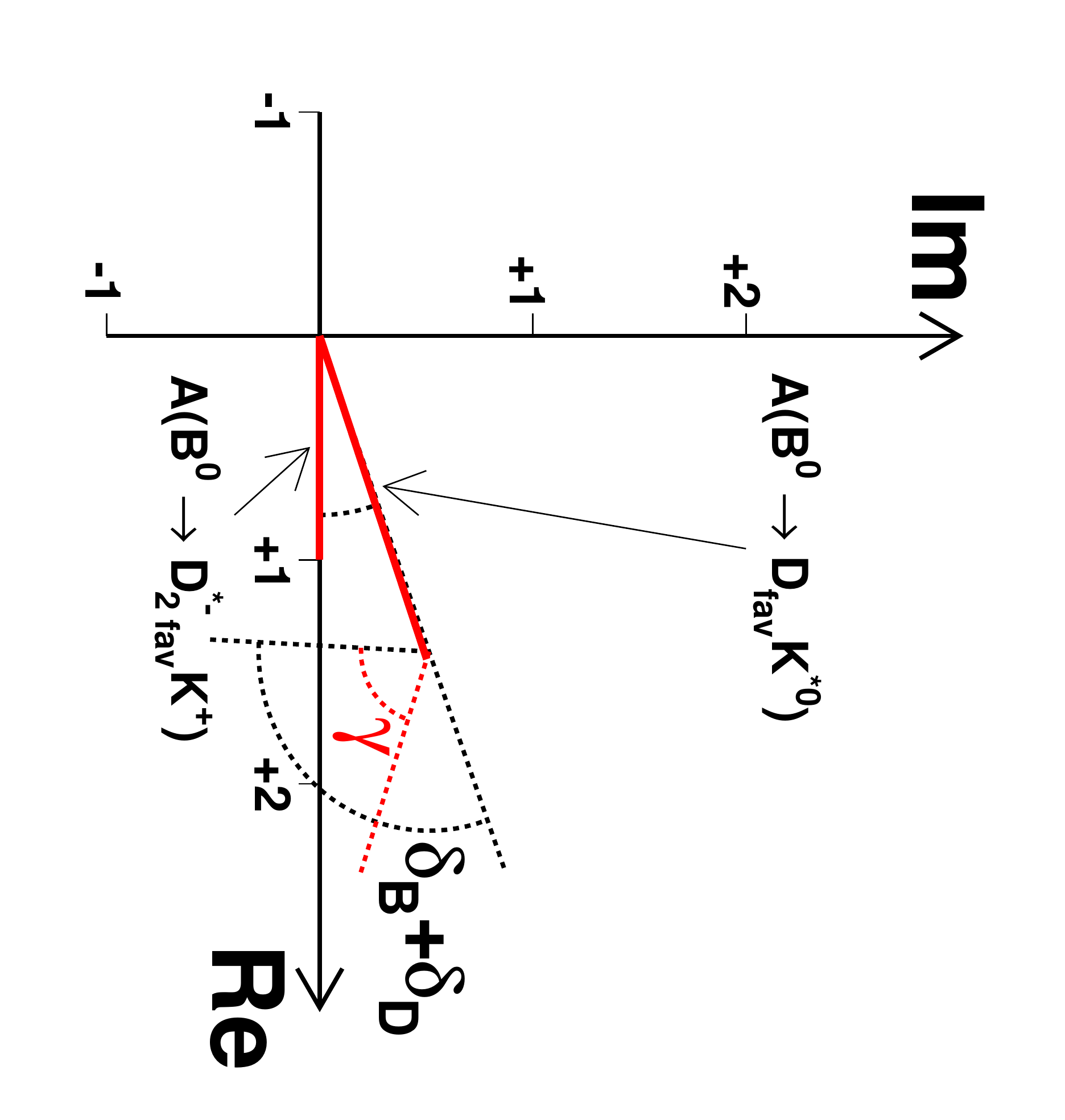}
  \hspace{8mm}
  \includegraphics[height=0.82\columnwidth,angle=90]{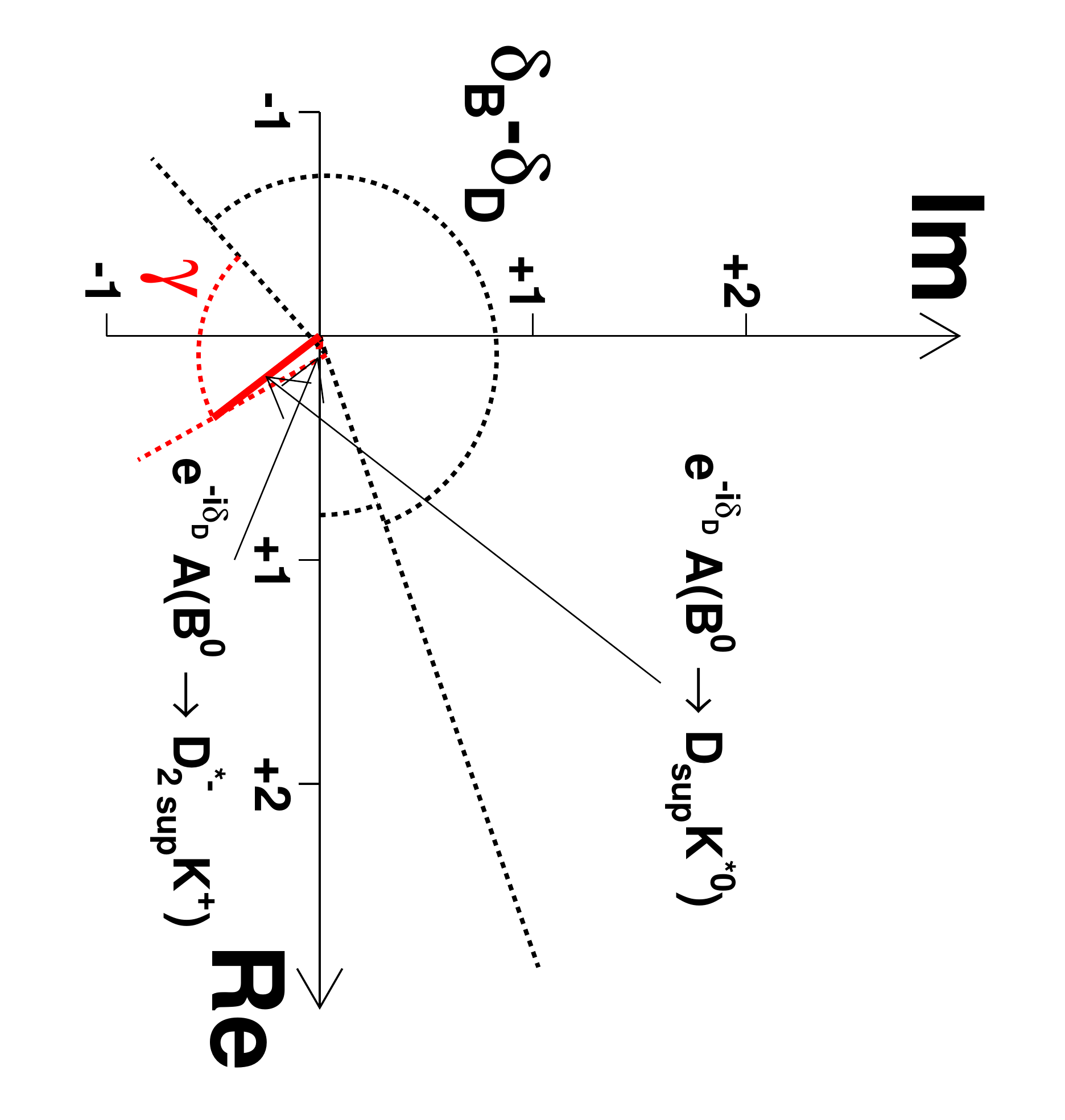}
  \caption{
    Argand diagrams illustrating the measurements of relative amplitudes 
    and phases from analysis of the Dalitz plots of 
    (left) $D_{\rm fav}K^+\pi^-$ and (right) $D_{\rm sup}K^+\pi^-$.
    Note that in the latter the amplitudes are rotated by $-\delta_D$ to
    maintain the convention of having the $D_2^{*-}K^+$ amplitude on the real
    axis.
    In these illustrative examples the following values are used:
    $\varrho = 1.5$, $\Delta = 20^\circ$, $\gamma = 75^\circ$,
    $\delta_B = 45^\circ$, $r_B = 0.4$,
    $\delta_D = -158^\circ$ and $r_D = 0.06$.
    \label{fig:argand2}
  }
\end{figure*}

The discussion above, and in Ref.~\cite{Gershon:2008pe}, considers that the
$D$ mesons used for normalisation are reconstructed in flavour-specific final
states.  However, as already mentioned, the favoured decay $D^0 \to K^-\pi^+$
is only approximately flavour-specific.  Furthermore, since the
doubly-Cabibbo-suppressed decays are used to great benefit in the
quasi-two-body analysis, it is reasonable to ask if they can also be included
in the Dalitz-plot analysis.  We therefore extend the method to include the
effects of the suppressed $D$-decay amplitudes.
We find
\begin{eqnarray}  
  \frac{A(B^0 \to D_{\rm fav}K^{*0})}{A(B^0 \to D_{2\,{\rm fav}}^{*-}K^+)} & = &
  \varrho e^{i\Delta} 
  \left( 1 + r_B r_D e^{i(\delta_B + \delta_D + \gamma)} \right)\, , \nonumber
  \\
  \label{eq:favAmpRatio} \\
  \frac{A(B^0 \to D_{\rm sup}K^{*0})}{A(B^0 \to D_{2\,{\rm sup}}^{*-}K^+)} & = &
  \varrho e^{i\Delta} 
  \left( 1 + \frac{r_B}{r_D} e^{i(\delta_B - \delta_D + \gamma)} \right)\, ,
  \nonumber \\
  \label{eq:supAmpRatio}
\end{eqnarray}
while the expression for the amplitude ratio in the Dalitz plot with the
$CP$-even $D$ meson is unchanged from Eq.~\ref{eq:cpAmpRatio}.
These amplitudes are illustrated in Fig.~\ref{fig:argand2}.
As before, the equivalent expressions for the charge-conjugate
$\overline{B}{}^0$ decays are obtained with the substitution 
$\gamma \longrightarrow -\gamma$.

As one would expect, the amplitudes obtained from the $D_{\rm fav}K^+\pi^-$
Dalitz plot (Fig.~\ref{fig:argand2}(left)) are hardly distinguishable from
those for the idealised flavour-specific $D$ decays.
As discussed in Ref.~\cite{Gershon:2008pe}, if the suppressed amplitudes are
neglected, it will lead to only a small bias in the extraction of $\gamma$.

The amplitudes obtained from the $D_{\rm sup}K^+\pi^-$ Dalitz plot
(Fig.~\ref{fig:argand2}(right)) are markedly different from those of the other
Dalitz plots, and this new information can in principle be used to constrain
$\gamma$.  However, in this case the $D_2^*K$ contribution is no longer
suitable as a reference amplitude, due to its small size, and the measurement
of the relative phase between $D_2^*K$ and $DK^*$ amplitudes would be expected
to have a large uncertainty. 
Nonetheless, it should be possible to obtain information about the relative
magnitude of these amplitudes, which will provide sensitivity to $\gamma$.
Note that, in contrast to the quasi-two-body analysis, it is not necessary to
use external constraints on $r_D$ and $\delta_D$ in the Dalitz-plot analysis.

\section{\boldmath $B \to DK\pi$ Dalitz Plot Model}
\label{sec:model}

We construct a model for $B \to DK\pi$ Dalitz plot distributions using the
isobar formalism, in which the total amplitude is written as the coherent sum
of contributions from resonant and nonresonant terms:
\begin{eqnarray}
  \label{eq:isobar}
  {\cal M}(\vec{x}) & = & a_{\rm nr} e^{i\phi_{\rm nr}} + \\
  & & \hspace{-5mm} \sum_r a_r e^{i\phi_r} F_{B\to Rb(\vec{x})}F_{R\to
    d_1d_2(\vec{x})} BW_r(\vec{x})S_r(\vec{x}). \nonumber
\end{eqnarray}
In Eq.~\ref{eq:isobar},
$\vec{x} = (m^2_{K\pi},m^2_{D\pi})$ represents the position in the Dalitz plot,
$a e^{i\phi}$ describes the complex amplitude for each component,
the $F$ terms denote vertex form factors,
$BW$ the resonance propagator
and $S$ the Lorentz invariant spin factor.
We use Blatt-Weisskopf barrier form factors~\cite{Blatt},
and use relativistic Breit-Wigner lineshapes to describe the propagators.
We use the Zemach formalism~\cite{Zemach:1963bc,Zemach:1968zz} for the spin
factors. 
We assume that the nonresonant contribution is constant across the
phase space.  
This is a sufficient approximation for the study at hand, even though a more
complicated description is likely to be necessary to fit real data.
All amplitudes are evaluated using the {\tt qft++} package~\cite{Williams:2008wu}.

We develop a model for the $B^0 \to D K^+\pi^-$ Dalitz plot distribution based
on the following results from Ref.~\cite{Aubert:2005yt}:
\begin{eqnarray}
  {\cal B}(B^0 \to D K^+\pi^-) \hspace{5mm} & \hspace{-10mm} = & \hspace{-5mm} (88 \pm 15 \pm 9) \times 10^{-6}, \nonumber \\
  {\cal B}(B^0 \to D K^{*0}(892)[K^+\pi^-]) & = & (38 \pm 6 \pm 4) \times 10^{-6}, \nonumber \\
  {\cal B}(B^0 \to D^{*-}_2(2460)[D\pi^-]K^+) & = & \nonumber \\
  & & \hspace{-10mm} (18.3 \pm 4.0 \pm 3.1) \times 10^{-6}, \nonumber \\
  {\cal B}(B^0 \to D K^+\pi^-)_{\rm nr} & = & (26 \pm 8 \pm 4) \times 10^{-6}. \nonumber 
\end{eqnarray}
The results for the resonant contributions were extracted using events in the
regions $|M_{K\pi} - M_{K^{*0}}| < 150$~MeV and $|M_{D\pi} - M_{D^{*-}_2}| <
75$~MeV, respectively.
The strengths of the $K^*(892)$ and $D^{*-}_2(2460)$ amplitudes in our model
were set by requiring that the fit fractions of these resonances in these
regions match the published results~\cite{Aubert:2005yt}.

Since we expect the Dalitz plot to contain contributions from other resonances
not considered in Ref.~\cite{Aubert:2005yt}, we add additional contributions
based on results from an analysis of the $B^0 \to D^\pm K_S^0 \pi^\mp$ Dalitz
plot~\cite{Aubert:2007qe}, taking isospin and colour-suppression factors into
account as appropriate. Some additional scaling of these resonance terms was
performed to provide a better match to the results published in  
Ref.~\cite{Aubert:2005yt}.
These experimental results provide information about the likely magnitude of
contributions from $K^*$ and $D^*$ resonances to the Dalitz plot.

Additionally, contributions from $D_s^*$-type resonances can in principle
contribute.  Their effect could be significant since they are mediated by 
$b \to u$ transitions which provide the sensitivity to $\gamma$.
The $D^*_{s2}(2573)$ and $D^*_{s1}(2700)$ states are known to decay to $DK$,
and the latter has been observed in $B$ decays~\cite{:2007aa}.
These are not included in our nominal model, but we consider their potential
effect among the model variations discussed below in
Section~\ref{sec:dalitz-results}.

\begin{table}
  \caption{
    Parameters of resonances used in the model~\cite{Amsler:2008zzb}.
  }
  \label{tab:resonance-params}
  \begin{tabular}{cccc}
    \hline
    Resonance & $J^P$ & Mass (MeV) & Width (MeV) \\
    \hline
    $K^*(892)$ &  $1^-$ & 896 & 51 \\
    $K^*_0(1430)$ & $0^+$ & 1412 & 294 \\
    $K^*_2(1430)$ & $2^+$ & 1432 & 109 \\
    $K^*(1680)$ & $1^-$ & 1717 & 322 \\
    $D^*_0(2400)$ & $0^+$ & 2403 & 283 \\
    $D^*_2(2460)$ & $2^+$ & 2459 & 25 \\
    $D^*_{s1}(2700)$ & $1^-$ & 2690 & 110 \\
    \hline
  \end{tabular}
\end{table}

\begin{figure}[!htb]
  \includegraphics[width=0.48\textwidth]{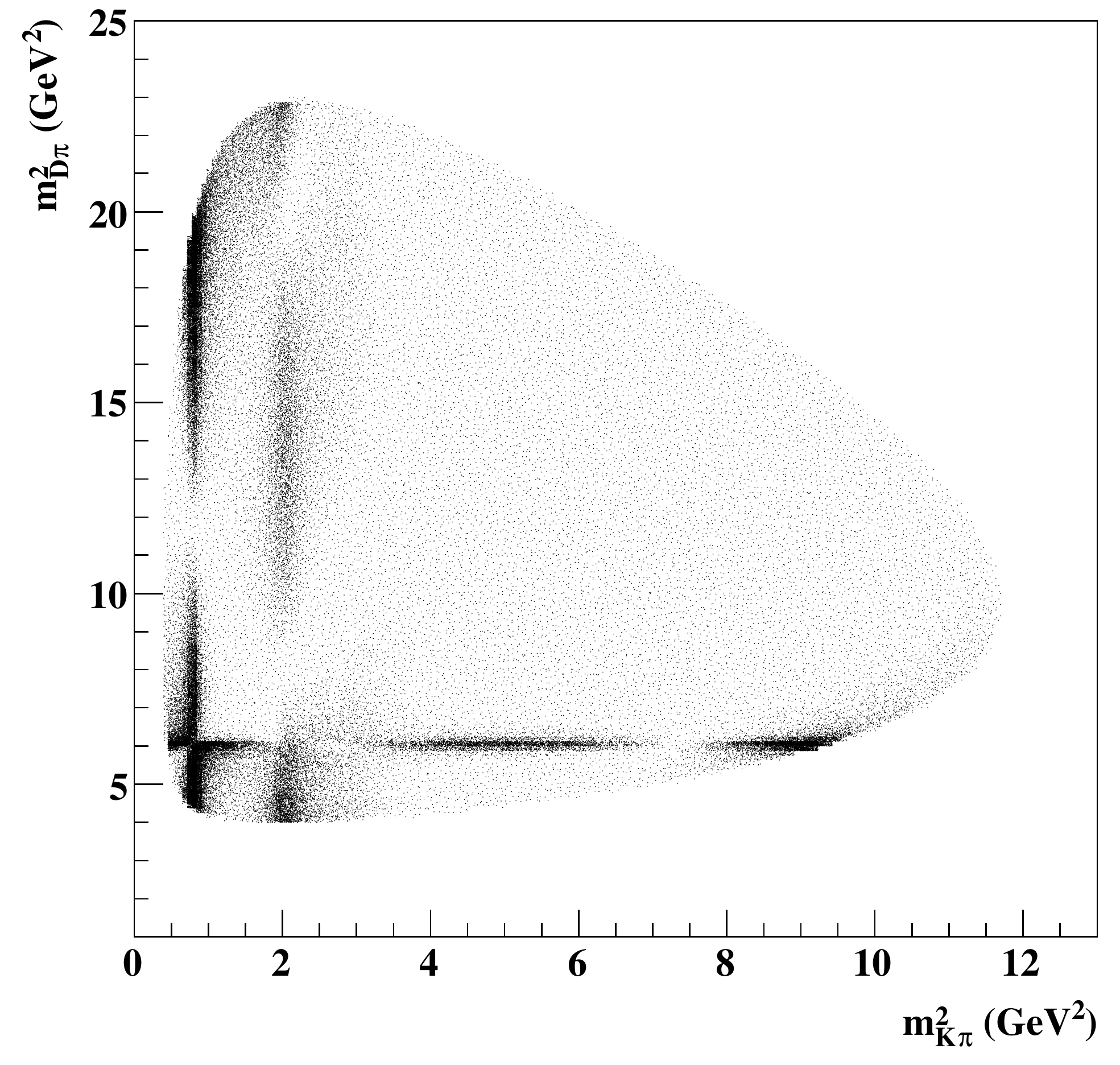}
  \caption{
    High-statistics $B\to D_{\rm fav}K\pi$ Dalitz plot distribution generated with 
    our nominal model.
    Structures due to $DK^{*0}(892)$, $DK^{*0}_2(1430)$ and $D^{*-}_2(2460)K^+$,
    as well as the nonresonant contribution, are clearly apparent.
  }
  \label{fig:model}
\end{figure}

Table~\ref{tab:resonance-params} summarises the parameters of the resonances
used in our analysis.
Table~\ref{tab:model} gives the fit fractions and relative phases of the
various intermediate states that contribute to our $B\to DK\pi$ Dalitz plot
model.  The parameters given are appropriate for the case that the neutral $D$
meson decays to a favoured, quasi-flavour-specific final state 
(namely, $B^0 \to DK^+\pi^-$, $D \to K^+ \pi^-$ and charge conjugate).
An example Dalitz-plot distribution generated from this model is shown in
Fig.~\ref{fig:model}. 

We derive the Dalitz-plot distributions for other $D$-decay final states using
the equations presented in Section~\ref{sec:method} and initially taking 
$\gamma = 60^\circ$.
We consider several different possible values of $r_B$ and $\delta_B$ for the
$DK^{*0}(892)$ amplitudes, 
while for the less significant contributions from other $DK^*$ channels we 
simply set $r_B = 0.4$ and $\delta_B = 0^{\circ}$.
We note that our nominal value of $r_B = 0.4$ is at the upper end of the
experimentally allowed range for this parameter ({\it e.g.} the UTfit
Collaboration gives $r_B(DK^{*0}(892)) \in \left[0.081,0.397\right]$ at
$95\%$ confidence level~\cite{Bona:2006ah}), but it is convenient to use this
value to allow comparison with previous studies.

\begin{table}
  \caption{
    Summary of the $B\to D_{\rm fav} K\pi$ Dalitz plot model.
  }
  \label{tab:model}
  \begin{tabular}{ccc}
    \hline
    Intermediate state & Fit fraction & Phase ($^\circ$) \\
    \hline
    $DK^{*0}(892)$     & 0.46     & 0   \\
    $DK_0^{*0}(1430)$   & 0.0001  & 284 \\
    $DK_2^{*0}(1430)$  & 0.12     & 221 \\
    $DK^{*0}(1680)$    & 0.02     & 128 \\
    $D_2^{*-}(2460)K^+$ & 0.34    & 325 \\
    $D_0^{-}(2400)K^+$  & 0.02    & 267 \\
    nonresonant $DK\pi$ & 0.06   &  140\\
    \hline
  \end{tabular}
\end{table}

We use our Dalitz-plot model to generate ensembles of event samples 
corresponding to the different $D$ meson final states.  The numbers of events
are based on expectations for one year of nominal luminosity at LHCb 
($2 \ {\rm fb}^{-1}$)~\cite{Akiba:2008zz}, 
assuming that the efficiency does not vary across the Dalitz plot. 
The exact numbers in each sample vary as functions of the input values of the
parameters (particularly $\delta_B$ and $\gamma$), but typically are around 
7300 for $D_{\rm fav}K\pi$, 700 for $D_{\rm sup}K\pi$ and 600 for $D_+K\pi$
(for $B^0$ and $\overline{B}{}^0$ decays combined).

\section{Results with the Quasi-two-body Approach}
\label{sec:q2b-results}

We first study the sensitivity to $\gamma$ in the quasi-two-body approach.  
We take the six Dalitz-plot distributions ($D_{\rm fav}K\pi$, $D_{\rm sup}K\pi$
and $D_+K\pi$ all for both $\overline{B}{}^0$ and $B^0$) generated as
described in the previous section, and apply selection requirements to select
the $DK^*$ dominated region. 
We then perform $\chi^2$ minimisation to fit the yields in each sample to
determine five parameters:  
$\gamma, r_S, \delta_S, \kappa$ and an overall normalisation.
The parameters $r_D$ and $\delta_D$ are fixed to their measured values 
($0.0616$ and $-158^{\circ}$, respectively).
For each pseudo-experiment we perform ten fits with initial values of the
parameters randomised in the ranges $\gamma \in \left[ 0, 2\pi \right)$, 
$r_B \in \left[ 0, 1 \right)$, $\delta_B \in \left[ 0, 2\pi \right)$,
$\kappa \in \left[ 0, 1 \right)$, and we take the results of the fit with the
smallest $\chi^2$.

The selection requirements for the $K^*$ are an interesting sub-topic worthy of
some discussion. 
The width of the $K\pi$ invariant-mass window around the nominal $K^*$ 
mass affects the size of the event samples. Increasing the width yields
greater statistics so that one would na\"ively expect a reduction in the
statistical uncertainty on $\gamma$. 
Unfortunately, increasing the width of the $K\pi$ invariant-mass window
also leads to an increase in the dilution from the non-$DK^*$ component of
the amplitude, {\em i.e.} a decrease in $\kappa$, which tends to decrease the
sensitivity to $\gamma$. 
Hence, it is important to optimise the $K^*$ selection requirements.

One can increase $\kappa$ by introducing a $D_2^*(2460)$ 
veto ({\em e.g.} rejecting events that satisfy 
${\left| M_{D\pi} - M_{D_2^*} \right| < 75 \ {\rm MeV}}$). 
This veto has a rather minimal impact on the statistics. 
In our nominal model with $\delta_B = 180^{\circ}$, choosing to use 
$\left| M_{K\pi} - M_{K^*(892)} \right| < 150 \ {\rm MeV}$ yields 
${\kappa = 0.93}$.  Applying the $D_2^*(2460)$ veto above increases this to
${\kappa = 0.97}$. Decreasing the $K\pi$ invariant-mass window to 
$\left| M_{K\pi} - M_{K^*(892)} \right| < 50 \ {\rm MeV}$ yields
${\kappa = 0.99}$; however, with the limited statistics in our 
pseudo-experiments,
this decrease in dilution is counter-balanced by the loss of statistics. 
Experiments with higher statistics may benefit by using a
tighter $K\pi$ invariant-mass window. 
In the results below, we proceed using the $D_2^*(2460)$ veto discussed above
and requiring the $K\pi$ invariant mass to be within 150~MeV of the nominal
$K^*$ mass.

\begin{figure*}[t]
  $\delta_B = 0^{\circ}$ \hspace{1.5in}$\delta_B = 45^{\circ}$ \hspace{1.5in} $\delta_B = 90^{\circ}$\\
  \includegraphics[width=0.3\textwidth]{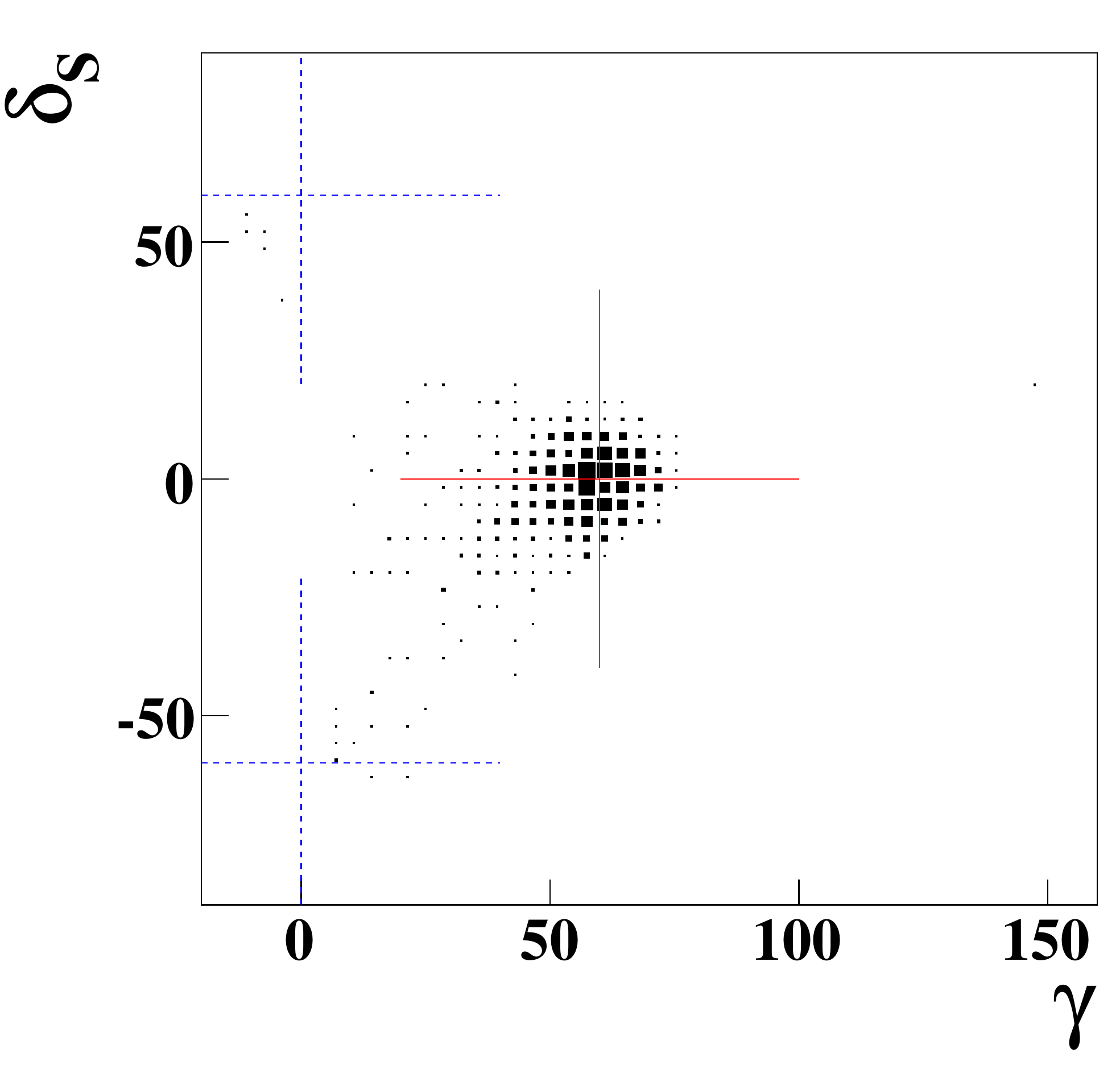}
  \includegraphics[width=0.3\textwidth]{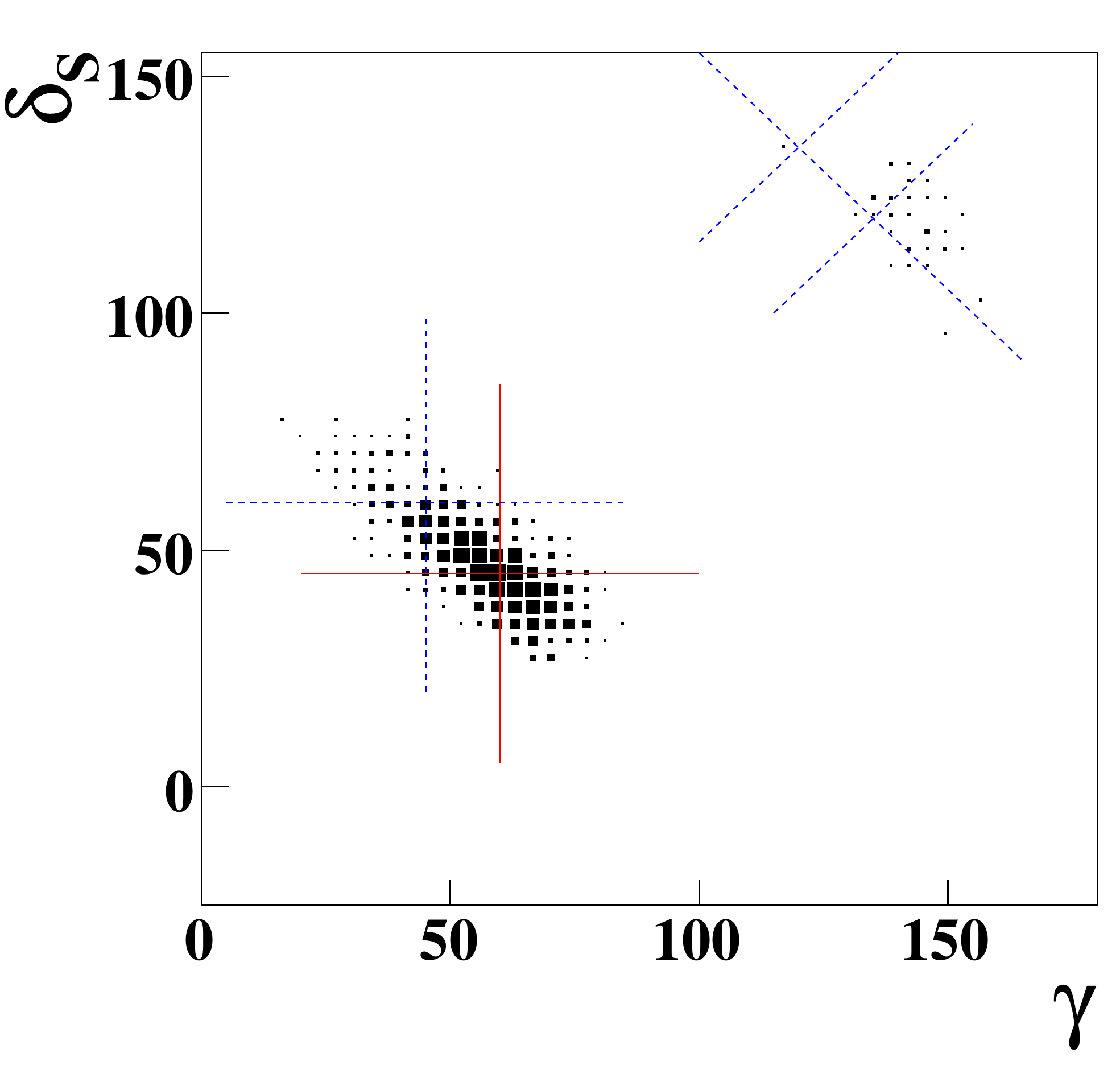}
  \includegraphics[width=0.3\textwidth]{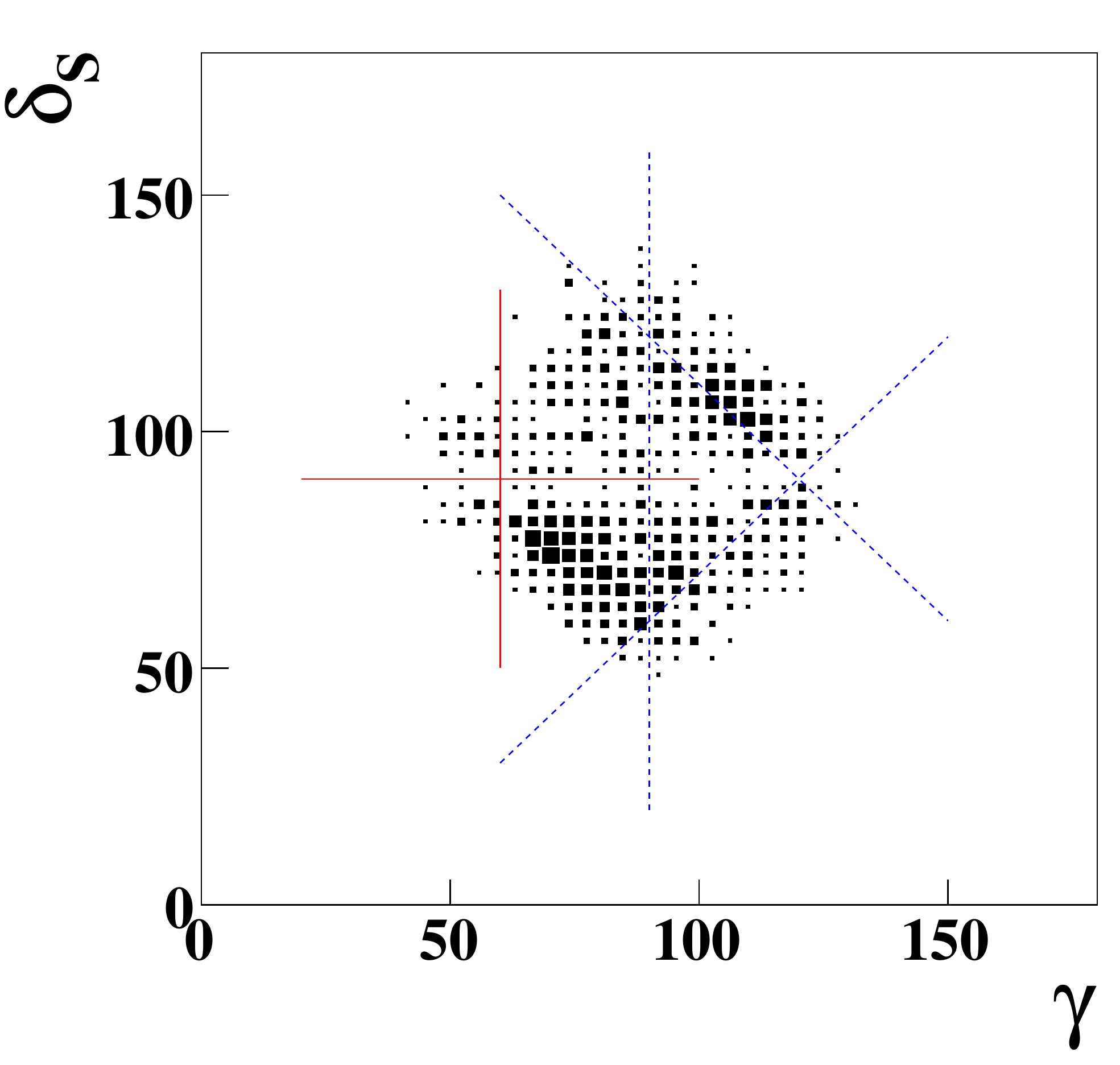}\\
  $\delta_B = 135^{\circ}$ \hspace{1.5in} $\delta_B = 180^{\circ}$\\
  \includegraphics[width=0.3\textwidth]{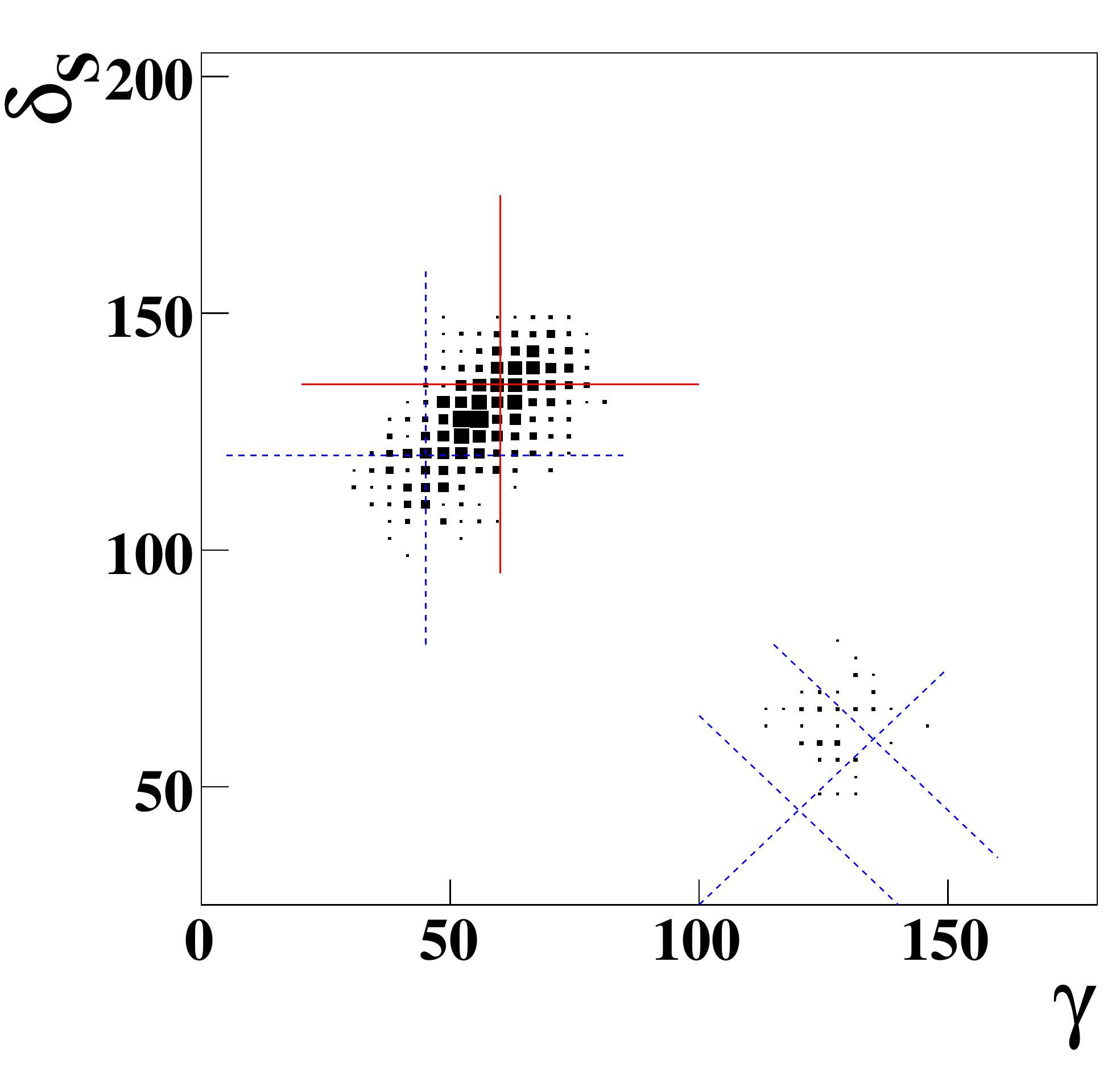}
  \includegraphics[width=0.3\textwidth]{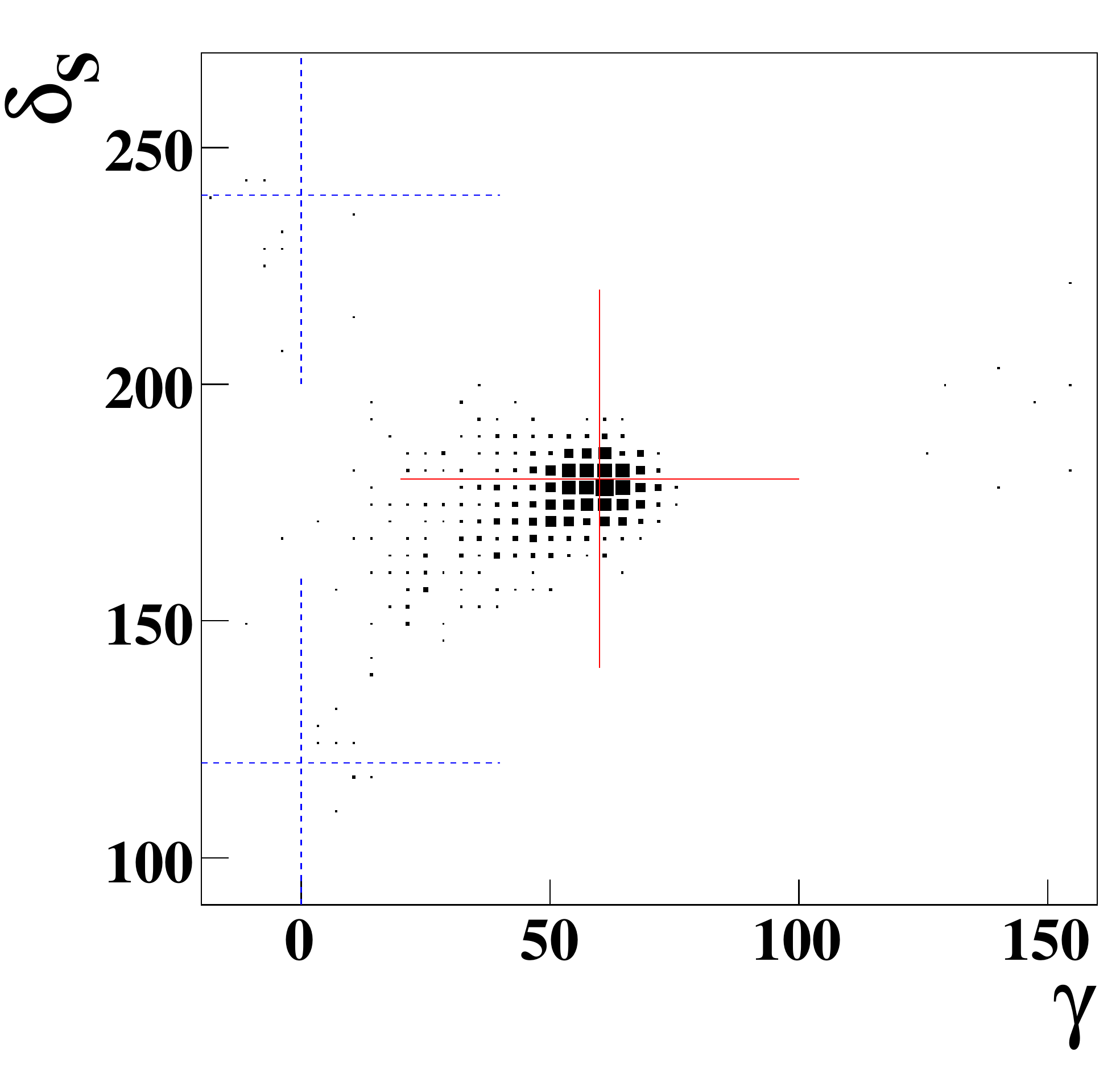}
  \caption{
    Distributions of fitted values of $\gamma$ in the quasi-two-body approach,
    with different input values of $\delta_B$.
    These results are obtained with $\gamma = 60^\circ$ and $r_B = 0.4$.
    The location of each of the generated solutions for $\kappa = 1$,
    {\em i.e.} $\delta_S = \delta_B$, is given by a red solid cross.
    The locations of each of the ambiguous solutions obtained for $\kappa = 1$
    and $r_D = 0$ are given by the intersections of the blue dashed lines.
  }
  \label{fig:q2b1}
\end{figure*}

Results from the pseudo-experiments for various different input values of
$\delta_B$ are shown in Fig.~\ref{fig:q2b1}.
It is apparent that there is a strong correlation
between the fitted values of $\gamma$ and $\delta_S$, and moreover that there
are ambiguities in the solution that are particularly pronounced for values of
$\delta_B$ near $90^\circ$. 
The locations of each of the ambiguous solutions for the case where $r_D = 0$ 
are also shown in Fig.~\ref{fig:q2b1}. The ambiguities in the solutions 
found by fitting our data are close to these values.
Since $r_D$ is small (our data were generated with $r_D = 0.0616$), 
statistical fluctuations can lead to the best $\chi^2$ value being obtained 
near one of the ambiguous solutions.
The inability to resolve such ambiguities is a limitation of the quasi-two-body
method.

As a consequence of these ambiguities, the distributions of the fitted values
of $\gamma$ are not Gaussian.
Therefore, in Tab.~\ref{tab:q2b1}, we report both the mean and rms of each
distribution as well as the corresponding values obtained by fitting the
data in the regions near the generated values of $\gamma$ to Gaussian 
lineshapes, {\em i.e.} ignoring the ambiguities in the solution. 
For $\delta_B =  90^{\circ}$, it is not possible to separate the correct
solutions from the ambiguous solution, so we have not performed a Gaussian fit
to the distributions for this sample.

\begin{figure*}[!htb]
  \includegraphics[width=0.3\textwidth]{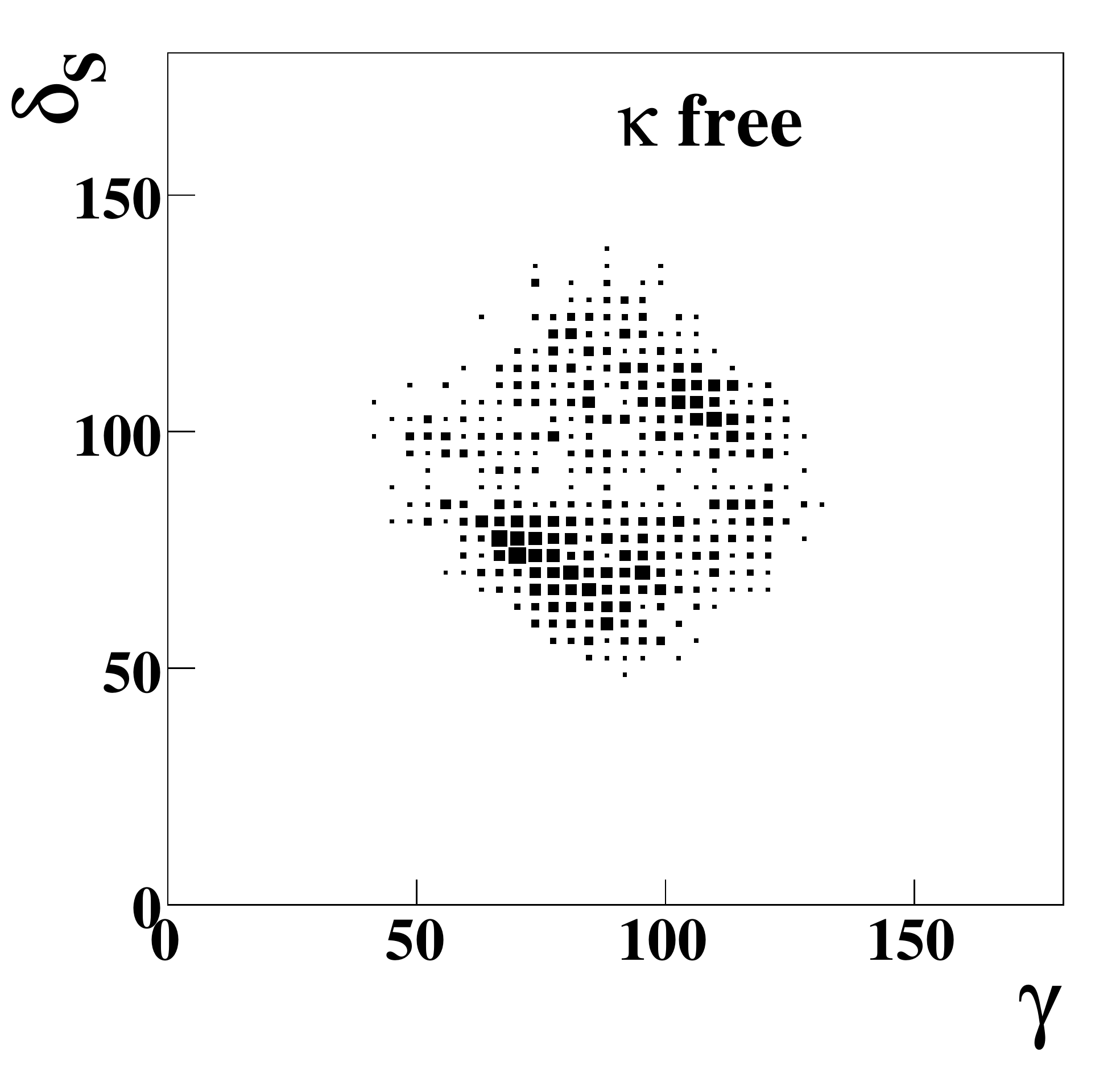}
  \hspace{0.05\textwidth}
  \includegraphics[width=0.3\textwidth]{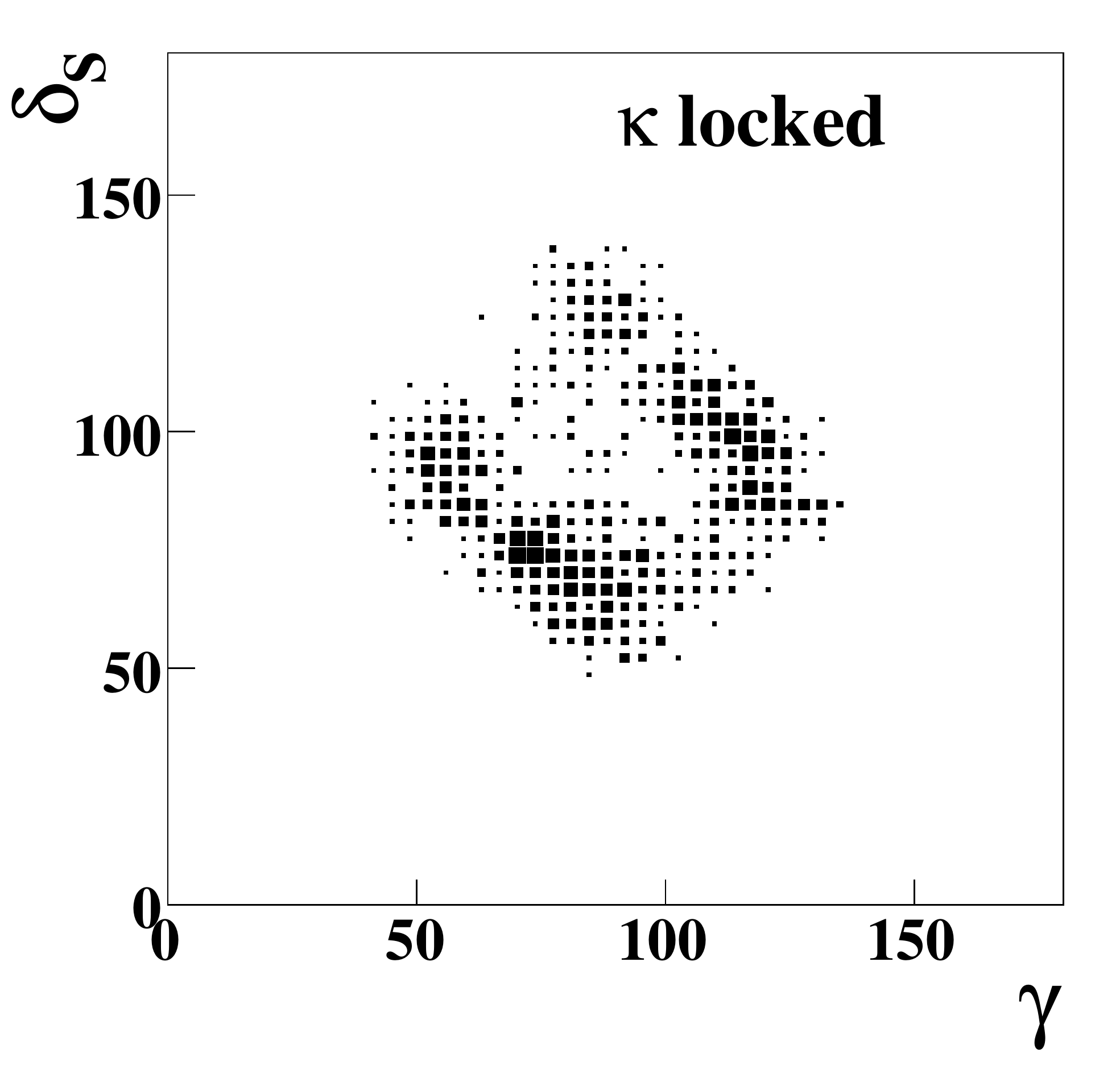}\\
  \caption{
    Distributions of fitted values of $\gamma$ in the quasi-two-body approach
    with $\kappa$ free (left) and fixed (right).  
    These results are obtained with 
    $\gamma = 60^{\circ}$, $\delta_B = 90^{\circ}$ and $r_B = 0.4$.
  }
  \label{fig:gamma-klock}
\end{figure*}

We have also considered an alternative approach to the fits, in which the
value of $\kappa$ is fixed.  Although $\kappa$ can only be measured from
analysis of $B \to DK\pi$ Dalitz plots (so that if this measurement can be
performed, the Dalitz plot analysis will most likely be feasible), it is
conceivable that $\kappa$ may be determined from theory, or from a different
experiment.  We therefore repeat the fits with $\kappa$ constrained to the
values obtained directly from our model.  This approach leads to an
improvement in the resolution on $\gamma$ near some solutions 
({\em e.g.}, near the true solution for $\delta_B = 0^{\circ}$),
but does not remove the ambiguities (see Fig.~\ref{fig:gamma-klock}).
The results obtained by fitting the regions near the generated values of
$\gamma$ to Gaussian lineshapes can be found in Tab.~\ref{tab:q2b1}.
This provides a comparison with Ref.~\cite{Akiba:2008zz},
where the same approach was used to estimate the sensitivity to $\gamma$ using
the quasi-two-body approach on one nominal year's data from LHCb. 
Despite several important differences between our toy study and the detailed
sensitivity study of Ref.~\cite{Akiba:2008zz},
the results are in very good agreement (see Tab.~\ref{tab:q2b1}).
This provides confidence in the absolute scale of our sensitivity estimates.

\begin{table}[!htb]
  \caption{
    Results for $\gamma$ obtained from the quasi-two-body analysis, 
    with different input values of $\delta_B$ and a comparison with the
    results of Ref.~\cite{Akiba:2008zz}.
    These results are obtained with $\gamma = 60^\circ$ and $r_B = 0.4$.  
    See text for details.
  }
  \label{tab:q2b1}
  \begin{tabular}{c|cc|cc|cc|c}
    \hline
    & \multicolumn{7}{c}{$\gamma$ ($^{\circ}$)} \\
    & \multicolumn{4}{c|}{$\kappa$ free} 
    & \multicolumn{2}{c|}{$\kappa$ locked} & Ref.~\cite{Akiba:2008zz} \\
    & \multicolumn{2}{c|}{Distribution} 
    & \multicolumn{2}{c|}{Gaussian fit} 
    & \multicolumn{2}{c|}{Gaussian fit} \\
    & mean & rms & $\mu$ & $\sigma$ & $\mu$ & $\sigma$ & $\sigma$ \\
    \hline
    $\delta_B =   0^{\circ}$ & 55.5 & 14.6 & 57.7 & 7.3 & 60.1 & 5.9 & 6.2\\
    $\delta_B =  45^{\circ}$ & 60.2 & 19.6 & 57.4 & 10.6 & 57.8 & 10.3 & 10.8 \\
    $\delta_B =  90^{\circ}$ & 88.3 & 17.9 & $-$ & $-$ & $-$ & $-$ & 12.7\\
    $\delta_B = 135^{\circ}$ & 60.4 & 18.0 & 56.6 & 9.4 & 56.3 & 9.7 & 9.5 \\
    $\delta_B = 180^{\circ}$ & 55.1 & 19.2 & 56.6 & 7.4 & 59.7 & 5.5 & 5.2\\
    \hline
  \end{tabular}
\end{table}

The results obtained by varying $r_B$ in our model are given in 
Tab.~\ref{tab:q2b2}.
As is well known, the sensitivity to $\gamma$ reduces as $r_B$ becomes
smaller.  It is also clear that the effect of ambiguities becomes more
significant, since the discrepancy between the rms of the whole distributions
and the width of the Gaussian peak near the correct solution increases.

\begin{table}[!htb]
  \caption{
    Results for $\gamma$ obtained from the quasi-two-body analysis, 
    with different input values of $r_B$.
    These results are obtained with $\gamma = 60^\circ$, $\delta_B = 0^{\circ}$ 
    and $\kappa$ free.  
  }
  \label{tab:q2b2}
  \begin{tabular}{c|cc|cc}
    \hline
    & \multicolumn{4}{c}{$\gamma$ ($^{\circ}$)} \\
    & \multicolumn{2}{c|}{Distribution} 
    & \multicolumn{2}{c}{Gaussian fit} \\
    & mean & rms & $\mu$ & $\sigma$ \\
    \hline
    $r_B = 0.4$ & 55.5 & 14.6 & 57.7 & 7.3 \\
    $r_B = 0.3$ & 57.0 & 21.1 & 56.8 & 9.7 \\
    $r_B = 0.2$ & 58.9 & 29.3 & 56.9 & 11.8 \\
    \hline
  \end{tabular}
\end{table}

%

\section{Results with the Amplitude Analysis}
\label{sec:dalitz-results}

To study the sensitivity to $\gamma$ using the Dalitz-plot analysis, we
perform a simultaneous maximum likelihood fit to all six Dalitz-plot
distributions.
We initially fix $r_D$ and $\delta_D$ to their measured values, leaving 24
free parameters ($\gamma$, four parameters -- $\Delta$, $\varrho$, $r_B$ \&
$\delta_B$ -- for each $K^*$-type resonance and nonresonant contribution, and
two -- magnitude and phase -- for each $D^*$-type resonance, with one phase
fixed).
As before, for each pseudo-experiment we perform multiple fits with randomised
initial parameter values, and we take the results of the fit with the smallest
negative log-likelihood.

\begin{figure*}[!htb]
  \includegraphics[width=0.3\textwidth]{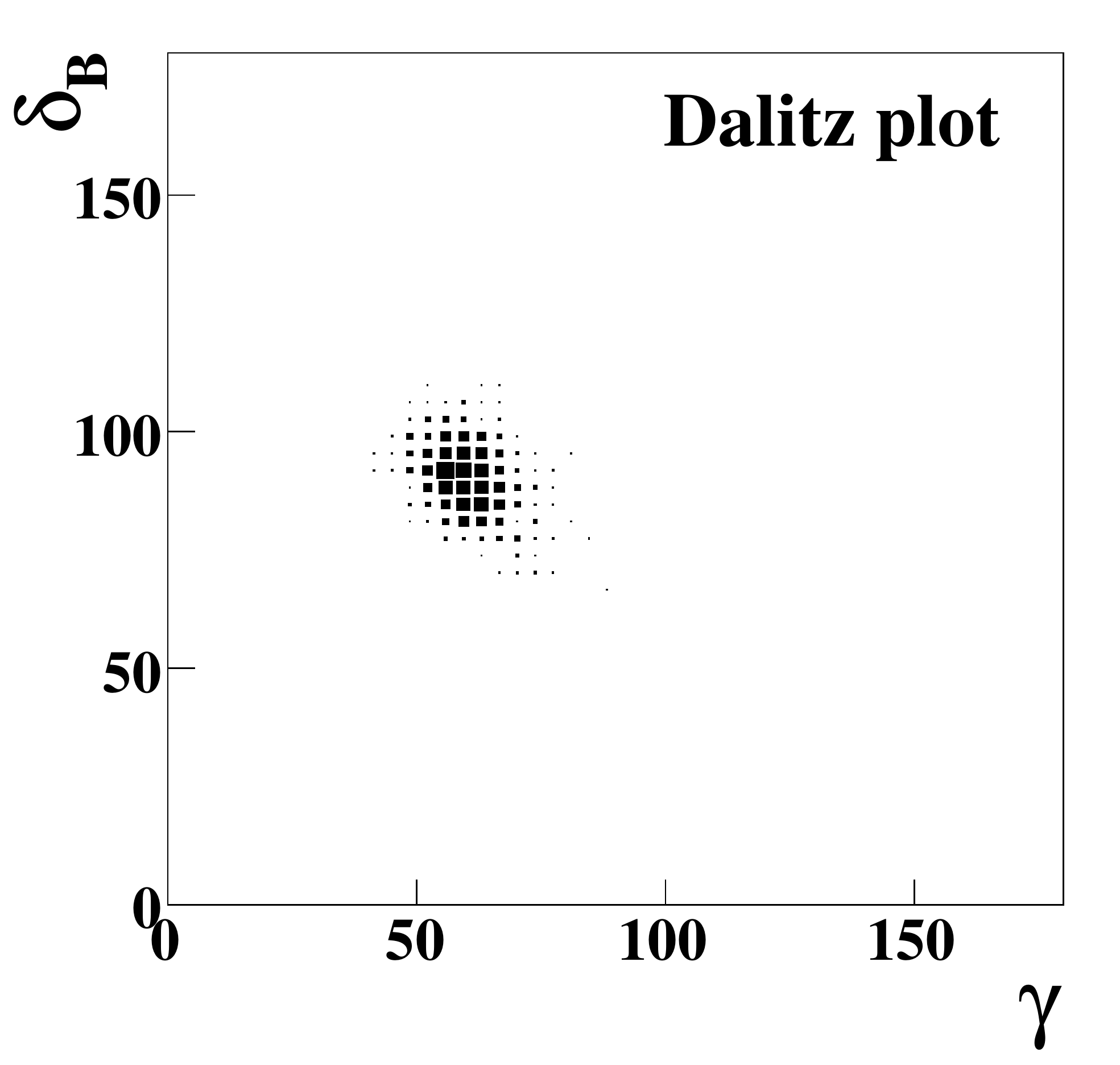}
  \hspace{0.05\textwidth}
  \includegraphics[width=0.3\textwidth]{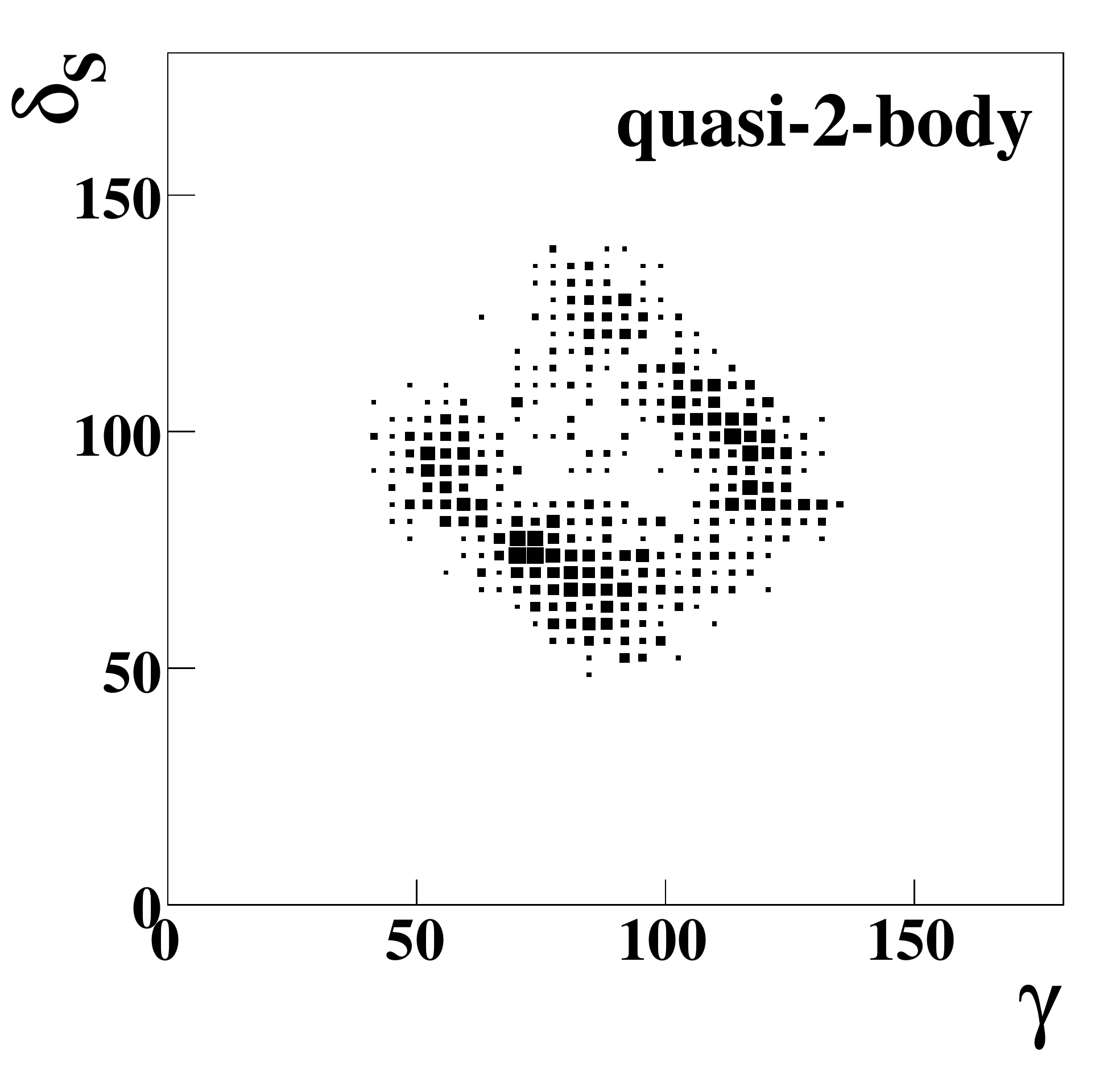}\\
  \caption{
    Distributions of fitted values of $\gamma$ in the amplitude (left) and 
    quasi-two-body (right) approaches.  These results are obtained with 
    $\gamma = 60^{\circ}$, $\delta_B = 90^{\circ}$ and $r_B = 0.4$.
  }
  \label{fig:gamma-comp}
\end{figure*}

In Section~\ref{sec:q2b-results}, we found that the quasi-two-body approach 
was, in many cases, unable to resolve ambiguities in the solution.
This was especially true for values of $\delta_B$ near $90^{\circ}$.  
Figure~\ref{fig:gamma-comp} shows the distributions of results 
from the pseudo-experiments generated with $\delta_B = 90^{\circ}$
for the amplitude and quasi-two-body approaches.  The additional information
utilised in the amplitude analysis is sufficient to remove the ambiguities
in the solution.

\begin{figure*}[!htb]
  \includegraphics[width=0.3\textwidth]{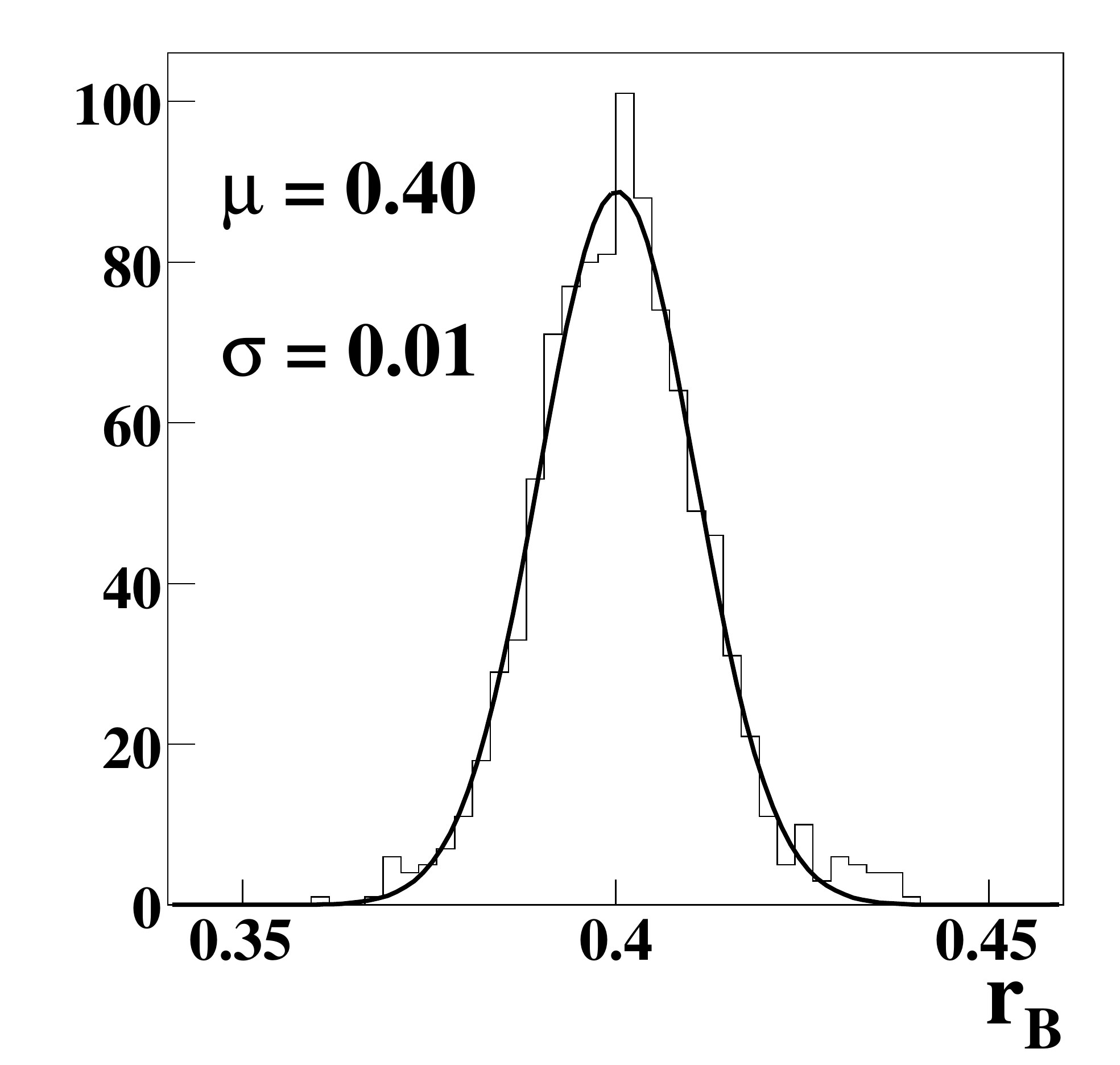}
  \includegraphics[width=0.3\textwidth]{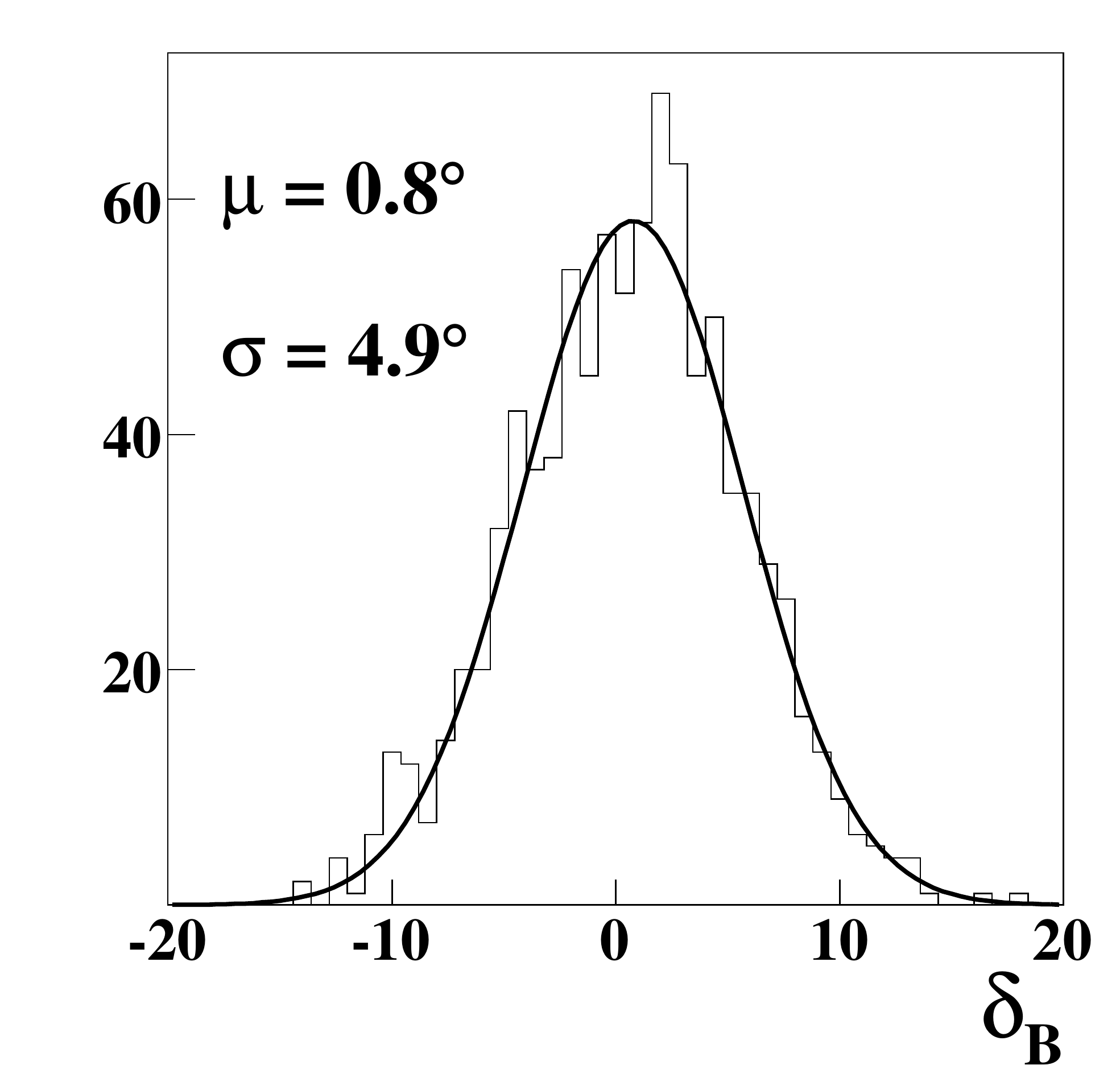}
  \includegraphics[width=0.3\textwidth]{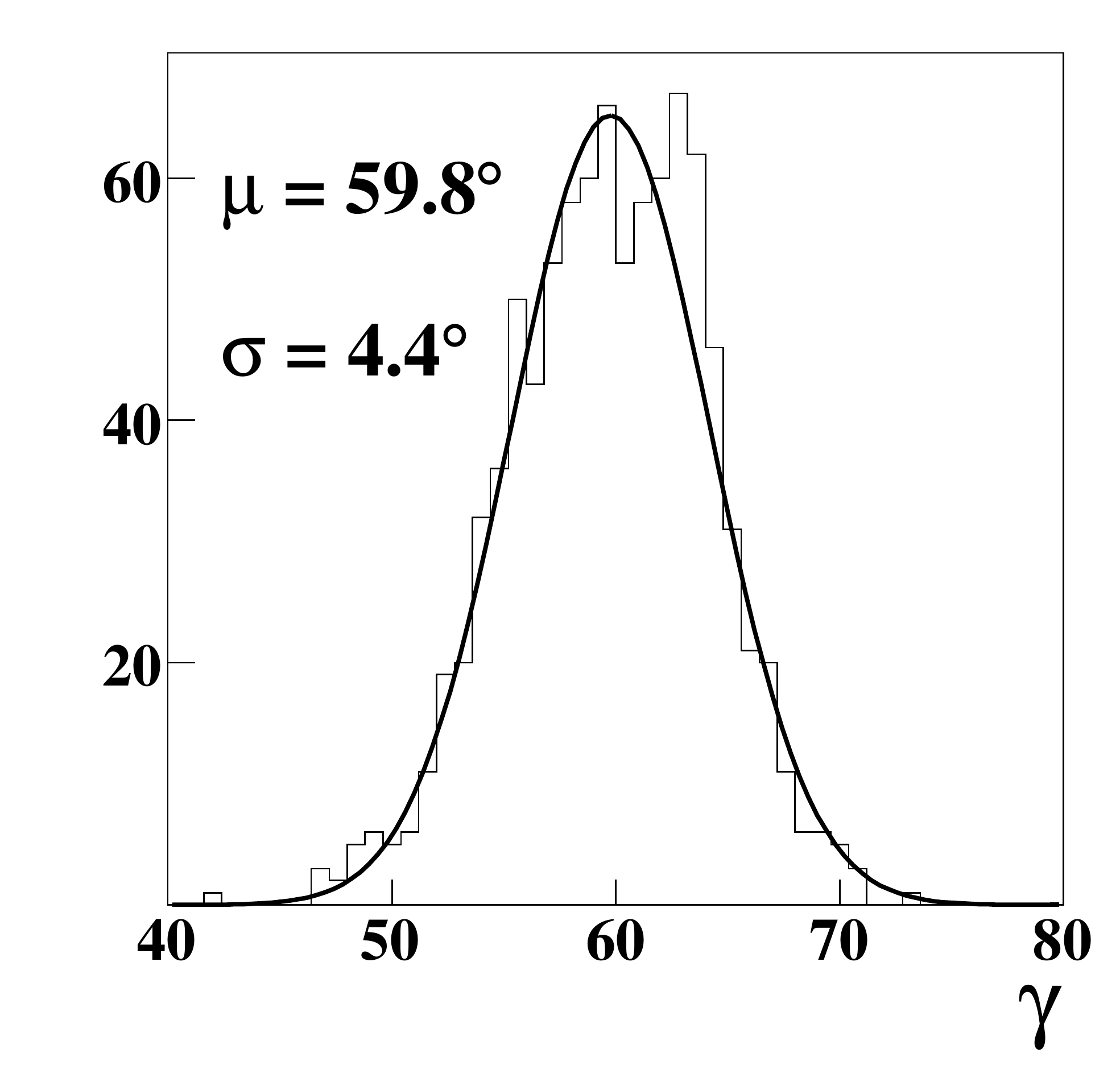}
  \caption{
    Distributions of fitted values of $r_B$ (left), $\delta_B$ (middle), and
    $\gamma$ (right) in the amplitude analysis.
    These results are obtained with 
    $\gamma = 60^{\circ}$, $\delta_B = 0^{\circ}$ and $r_B = 0.4$.
  }
  \label{fig:gamma-res}
\end{figure*}


The distributions of the fit results for the parameters $r_B(DK^*(892))$,
$\delta_B(DK^*(892))$ and $\gamma$ obtained from the pseudo-experiments 
generated with $\delta_B = 0^{\circ}$ are shown in Fig.~\ref{fig:gamma-res}.
We fit the distributions with Gaussian lineshapes, 
since all are consistent with this 
shape (this is true for all $\delta_B$ values), 
and report the means and widths in Tab.~\ref{tab:res-deltaB}.
The resolution on $\gamma$ varies between $4.2^{\circ}-5.8^{\circ}$, depending
on the value of $\delta_B$, in our nominal model.  
Thus, unlike for the quasi-two-body approach (Tab.~\ref{tab:q2b1}), 
the sensitivity to $\gamma$ is not strongly dependent on the value of 
$\delta_B$.

\begin{table}[!htb]
  \caption{
    Results for $r_B(DK^*(892))$, $\delta_B(DK^*(892))$ and $\gamma$
    from the Dalitz-plot analysis, 
    with different input values of $\delta_B$.
    These results are obtained with $\gamma = 60^\circ$ and $r_B = 0.4$.
  }
  \label{tab:res-deltaB}
  \begin{tabular}{c|cc|cc|cc}
    \hline
    & \multicolumn{2}{c|}{$r_B$} & \multicolumn{2}{c|}{$\delta_B$ ($^{\circ}$)}
    & \multicolumn{2}{c}{$\gamma$ ($^{\circ}$)} \\
    & $\mu$ & $\sigma$ & $\mu$ & $\sigma$ & $\mu$ & $\sigma$ \\
    \hline
    $\delta_B =   0^{\circ}$ & 0.40 & 0.01 &   0.8 & 4.9 & 59.8 & 4.4 \\
    $\delta_B =  45^{\circ}$ & 0.40 & 0.01 &  46.2 & 6.0 & 59.2 & 5.5 \\
    $\delta_B =  90^{\circ}$ & 0.40 & 0.01 &  90.2 & 6.4 & 59.8 & 5.7 \\
    $\delta_B = 135^{\circ}$ & 0.40 & 0.01 & 134.0 & 6.3 & 59.3 & 5.8 \\
    $\delta_B = 180^{\circ}$ & 0.40 & 0.01 & 179.8 & 4.2 & 59.7 & 4.2 \\
    \hline
  \end{tabular}
\end{table}

The physical values of $r_B$ and $\gamma$ may be different from those used in
our model.  
To study how this would affect our results, we vary the values of $r_B$ and 
$\gamma$ in our model and fit the data using the same procedure outlined above.
The results obtained considering $0.2 \leq r_B \leq 0.4$ are given in 
Tab.~\ref{tab:res-rB}.  
As expected, the sensitivity to $\gamma$ decreases as $r_B$ decreases;
however, even for $r_B=0.2$ the distribution of fit results is still 
approximately Gaussian, and the resolution on $\gamma$ is still $7^{\circ}$.
The results obtained considering $45^{\circ} \leq \gamma \leq 75^{\circ}$ are
given in Tab.~\ref{tab:res-gamma}.   
We conclude that the sensitivity to $\gamma$ does not depend strongly on the
value of $\gamma$ itself.

\begin{table}[!htb]
  \caption{
    Results for $r_B(DK^*(892))$, $\delta_B(DK^*(892))$ and $\gamma$
    from the Dalitz-plot analysis, 
    with different input values of $r_B$.
    These results are obtained with $\gamma = 60^\circ$ and $\delta_B = 0^{\circ}$.
  }
  \label{tab:res-rB}
  \begin{tabular}{c|cc|cc|cc}
    \hline
    & \multicolumn{2}{c|}{$r_B$} & \multicolumn{2}{c|}{$\delta_B$ ($^{\circ}$)}
    & \multicolumn{2}{c}{$\gamma$ ($^{\circ}$)} \\
    & $\mu$ & $\sigma$ & $\mu$ & $\sigma$ & $\mu$ & $\sigma$ \\
    \hline
    $r_B = 0.4$ & 0.40 & 0.01 & 0.8 & 4.9 & 59.8 & 4.4 \\
    $r_B = 0.3$ & 0.30 & 0.01 & 0.7 & 5.9 & 60.0 & 5.3 \\
    $r_B = 0.2$ & 0.20 & 0.01 & 0.6 & 7.0 & 59.5 & 7.0 \\
    \hline
  \end{tabular}
\end{table}

\begin{table}[!htb]
  \caption{
    Results for $r_B(DK^*(892))$, $\delta_B(DK^*(892))$ and $\gamma$
    from the Dalitz-plot analysis, 
    with different input values of $\gamma$.
    These results are obtained with $r_B = 0.4$ and $\delta_B = 0^{\circ}$.
  }
  \label{tab:res-gamma}
  \begin{tabular}{c|cc|cc|cc}
    \hline
    & \multicolumn{2}{c|}{$r_B$} & \multicolumn{2}{c|}{$\delta_B$ ($^{\circ}$)}
    & \multicolumn{2}{c}{$\gamma$ ($^{\circ}$)} \\
    & $\mu$ & $\sigma$ & $\mu$ & $\sigma$ & $\mu$ & $\sigma$ \\
    \hline
    $\gamma = 45^{\circ}$ & 0.40 & 0.01 & 1.2 & 6.0 & 44.6 & 5.2 \\
    $\gamma = 60^{\circ}$ & 0.40 & 0.01 & 0.8 & 4.9 & 59.8 & 4.4 \\
    $\gamma = 75^{\circ}$ & 0.40 & 0.01 & 0.4 & 4.5 & 74.8 & 4.0 \\
    \hline
  \end{tabular}
\end{table}

There are a number of parameters in our model which are not well
constrained by data; however, most of these are not expected to impact
significantly the sensitivity to $\gamma$.  
For example, the strong phase between the $D^{*-}_2(2460)K^+$ and
$\bar{D}K^{*0}(892)$ amplitudes, which we denote by $\Delta[DK^{*0}(892)]$,
could be very different from the value used in our model.  
To determine what effect the value of this parameter has on the extracted
value of $\gamma$, we vary $\Delta[DK^{*0}(892)]$ in our model over the 
physically allowed range, $[0,2\pi)$, and refit the data.
The fit results, given in Tab.~\ref{tab:res-Delta}, show that the impact on
the sensitivity to $\gamma$ is minimal.

\begin{table}[!htb]
  \caption{
    Results for $r_B(DK^*(892))$, $\delta_B(DK^*(892))$ and $\gamma$
    from the Dalitz-plot analysis, 
    with different input values of $\Delta[DK^{*0}(892)]$.
    These results are obtained with $r_B = 0.4$, $\delta_B = 0^{\circ}$ and 
    $\gamma = 60^{\circ}$.
  }
  \label{tab:res-Delta}
  \begin{tabular}{c|cc|cc|cc}
    \hline
    $\Delta[DK^{*0}(892)]$
    & \multicolumn{2}{c|}{$r_B$} & \multicolumn{2}{c|}{$\delta_B$ ($^{\circ}$)}
    & \multicolumn{2}{c}{$\gamma$ ($^{\circ}$)} \\
    & $\mu$ & $\sigma$ & $\mu$ & $\sigma$ & $\mu$ & $\sigma$ \\
    \hline
    $55^{\circ}$ & 0.40 & 0.01 & 0.5 & 4.9 & 59.7 & 4.1 \\
    $145^{\circ}$ & 0.40 & 0.01 & 0.4 & 4.9 & 59.7 & 4.3 \\
    $235^{\circ}$ & 0.40 & 0.01 & 0.0 & 5.0 & 59.5 & 4.0 \\
    $325^{\circ}$ & 0.40 & 0.01 & 0.8 & 4.9 & 59.8 & 4.4 \\
    \hline
  \end{tabular}
\end{table}

None of the parameters for the $DK^{*0}_0(1430)$, $DK^{*0}_2(1430)$ or 
$DK^{*0}(1680)$ amplitudes is well constrained by data.  
To see what effect the presence of $CP$ violation in these amplitudes has on 
the sensitivity to $\gamma$, we set
$r_B(DK^{*0}_0(1430)) = 0$, $r_B(DK^{*0}_2(1430)) = 0$ and 
$r_B(DK^{*0}(1680)) = 0$ in our model and refit the data.  
The results obtained for this model variation are given in 
Tab.~\ref{tab:res-other-rB}.  
There is no appreciable bias on the extracted value of $\gamma$, and the
resolution on $\gamma$ decreases by only $1.2^{\circ}$.
This is not surprising given the relatively small contributions to the total
$B\rightarrow DK\pi$ amplitude from $DK^{*0}_0(1430)$, $DK^{*0}_2(1430)$ and 
$DK^{*0}(1680)$ (see Tab.~\ref{tab:model}).
If, however, in real data these or any other $DK^*$-type amplitudes do
contribute strongly to $B\rightarrow DK\pi$ 
(and have sufficiently large $CP$ violation), then the sensitivity to $\gamma$
could be better than that obtained from our nominal model.
To further test how mismodelling of the Dalitz plot may affect the extracted
value of $\gamma$, we refit the data from our nominal model 
using only the $DK^*(892)$,
$D^-_2(2460)K$ and nonresonant amplitudes (fixing all others to zero).
The results are shown in Tab.~\ref{tab:res-messed-up}.  
Again, there is little impact on the extracted values, indicating that this
analysis may suffer much less model-related uncertainties than some other 
methods to extract $\gamma$~\cite{Poluektov:2006ia,Aubert:2008bd}.

\begin{table}[!htb]
  \caption{
    Results for $r_B(DK^*(892))$, $\delta_B(DK^*(892))$ and $\gamma$
    from the Dalitz-plot analysis, 
    with $r_B(DK^{*0}_0(1430)) = 0$, $r_B(DK^{*0}_2(1430)) = 0$ and 
    $r_B(DK^{*0}(1680)) = 0$.
    These results are obtained with $r_B = 0.4$, $\delta_B = 0^{\circ}$ and 
    $\gamma = 60^{\circ}$.
  }
  \label{tab:res-other-rB}
  \begin{tabular}{cc|cc|cc}
    \hline
    \multicolumn{2}{c|}{$r_B$} & \multicolumn{2}{c|}{$\delta_B$ ($^{\circ}$)}
    & \multicolumn{2}{c}{$\gamma$ ($^{\circ}$)} \\
    $\mu$ & $\sigma$ & $\mu$ & $\sigma$ & $\mu$ & $\sigma$ \\
    \hline
    0.40 & 0.01 & 0.1 & 5.2 & 59.0 & 5.6 \\
    \hline
  \end{tabular}
\end{table}

\begin{table}[!htb]
  \caption{
    Results for $r_B(DK^*(892))$, $\delta_B(DK^*(892))$ and $\gamma$
    from the Dalitz-plot analysis, 
    where the fit model contains only the $DK^*(892)$, $D^-_2(2460)K$ and 
    nonresonant amplitudes. 
    These results are obtained with $r_B = 0.4$, $\delta_B = 0^{\circ}$ and 
    $\gamma = 60^{\circ}$.
  }
  \label{tab:res-messed-up}
  \begin{tabular}{cc|cc|cc}
    \hline
    \multicolumn{2}{c|}{$r_B$} & \multicolumn{2}{c|}{$\delta_B$ ($^{\circ}$)}
    & \multicolumn{2}{c}{$\gamma$ ($^{\circ}$)} \\
    $\mu$ & $\sigma$ & $\mu$ & $\sigma$ & $\mu$ & $\sigma$ \\
    \hline
    0.40 & 0.01 & -1.0 & 4.4 & 60.2 & 4.5 \\
    \hline
  \end{tabular}
\end{table}

We have also tested how the addition of $D_{s}$-type resonances may affect the
analysis.  We add to the model a contribution from $D_{s1}^*(2700)$, the
magnitude of which is based on the assumption
\begin{equation}
  \frac{{\cal B}(B\to D_{sJ}\pi)}{{\cal B}(B\to D_{sJ}D)} \approx
  \frac{{\cal B}(B\to D_s^{(*)}\pi)}{{\cal B}(B\to D_s^{(*)}D)} \, .
\end{equation}
The right-hand side of this expression is found to be 
$(2.4 \pm 0.6) \times 10^{-3}$ for $D_s$ and 
$(3.3 \pm 0.9) \times 10^{-3}$ for $D_s^*$~\cite{Amsler:2008zzb}.
Using results from Ref.~\cite{:2007aa}, we estimate 
\begin{eqnarray}
  {\cal B}(B\to D_{s1}^*(2700)\pi)\times{\cal B}(D_{s1}^*(2700)\to DK) 
  & \approx & \\
  && \hspace{-50mm} 
  2.7 \times 10^{-3} \times 11.3 \times 10^{-4} \approx 3 \times 10^{-6} \, .
  \nonumber
\end{eqnarray}

We add a contribution from this amplitude, with an arbitrary phase, to the
model described in Tab.~\ref{tab:model}.
The results of fits to pseudo-experiments generated with this modified model
are summarised in Tab.~\ref{tab:res-dsstar}.
There is little or no improvement in the sensitivity to $\gamma$ compared to
the nominal model.
This can be understood since the corner of the Dalitz plot where the
$D_{s1}^*(2700)$ and $K^*(892)$ would interfere is excluded by kinematic
constraints (see Fig.~\ref{fig:model}).
Following this reasoning, larger effects from $D_{sJ}$-type resonances would
be possible if there were significant contributions to the Dalitz plot from
heavier $K^*$ and/or heavier $D_{sJ}$ resonances.

\begin{table}[h!]
 \caption{
   Results for $r_B(DK^*(892))$, $\delta_B(DK^*(892))$ and $\gamma$
   obtained from the Dalitz-plot analysis,
   where the $D_{s1}^*(2700)$ has been added to our nominal model.
   These results are for $r_B = 0.4$, $\delta_B = 0^{\circ}$ and
   $\gamma = 60^{\circ}$.
 }
 \label{tab:res-dsstar}
 \begin{tabular}{cc|cc|cc}
   \hline
   \multicolumn{2}{c|}{$r_B$} & \multicolumn{2}{c|}{$\delta_B$ ($^{\circ}$)}
   & \multicolumn{2}{c}{$\gamma$ ($^{\circ}$)} \\
   $\mu$ & $\sigma$ & $\mu$ & $\sigma$ & $\mu$ & $\sigma$ \\
   \hline
   0.40 & 0.01 & 0.8 & 5.0 & 59.9 & 4.3 \\
   \hline
 \end{tabular}
\end{table} 

\begin{figure}[!htb]
  \includegraphics[width=0.3\textwidth]{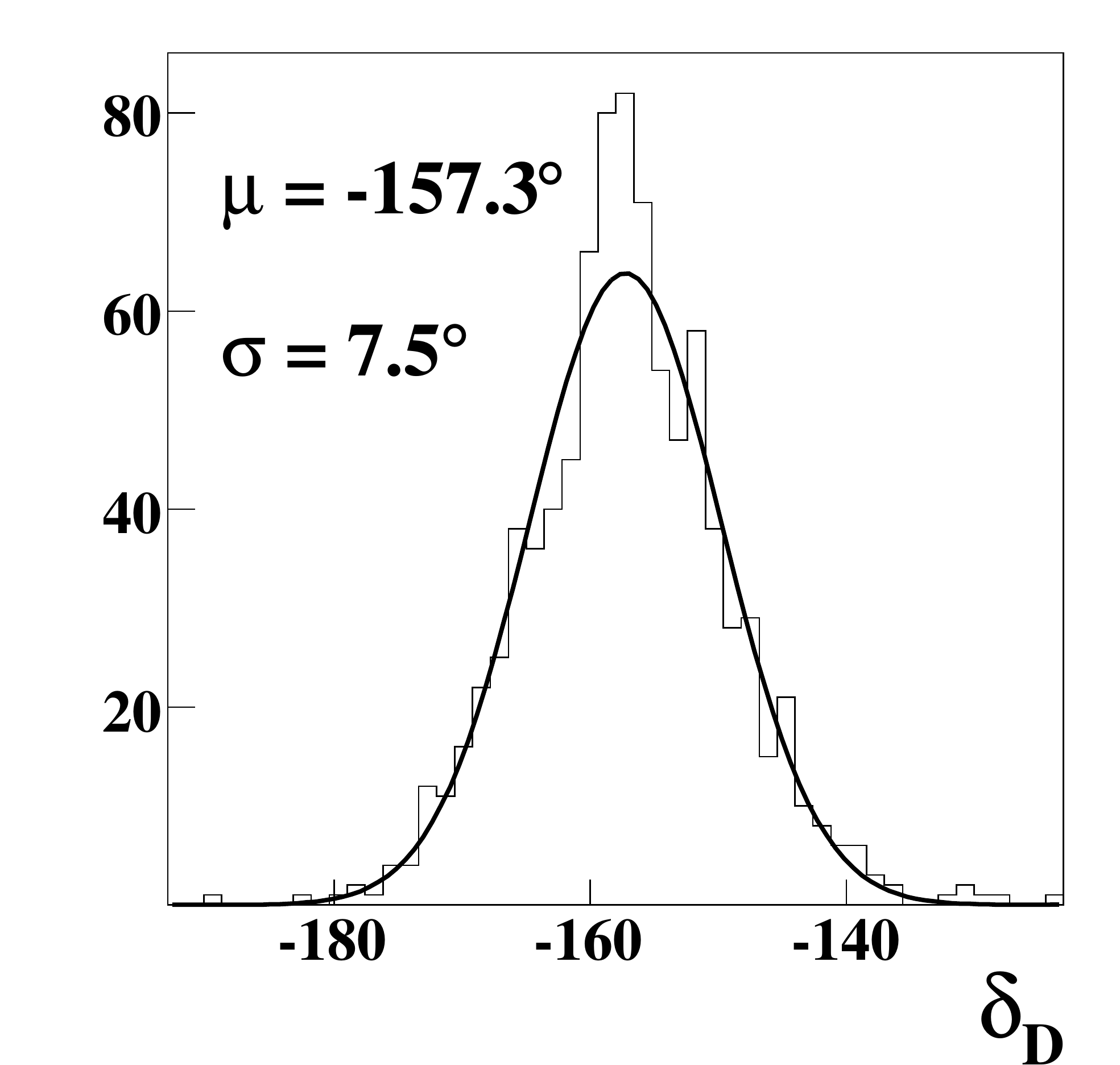}
  \caption{Distribution of fitted results for $\delta_D$ obtained in the 
    amplitude analysis when fitting with $\delta_D$ as a free parameter.
    The mean and width obtained by fitting the distribution to a Gaussian
    lineshape (solid line) are shown on the plot.  
    The generated value of $\delta_D$ is $-158^{\circ}$.
  }
  \label{fig:deltaD-res}
\end{figure}

In our nominal fit, we fix the values of $\delta_D$ and $r_D$.
However, the value of $\delta_D$ is currently not precisely measured:
$\delta_D = (-158 \pm 11)^\circ$~\cite{Rosner:2008fq,Asner:2008ft,Barberio:2008fa}.
To study what effect this parameter can have on our results, we
rerun the fits described above (on our nominal model) 
starting the value of $\delta_D$ randomly in the range $[0,2\pi)$.  
The results obtained for $\delta_D$ are shown in Fig.~\ref{fig:deltaD-res}.  
The resolution extracted for $\delta_D$ is $7.5^{\circ}$, which is less than
the uncertainty on current measurements.  
Fitting with $\delta_D$ free has no significant effect on our results for
$r_B(DK^*(892))$, $\delta_B(DK^*(892))$ and $\gamma$ (for all model values of
$\delta_B$).

\section{Experimental Effects}
\label{sec:effects}

The study reported in the previous section neglected a number of experimental
effects which will surely impact the sensitivity to $\gamma$ in any real-world
experiment.  
The effect most likely to have the largest impact on the resolution is the
presence of background events.  To study how this might affect the
sensitivity to $\gamma$, we generate background events by sampling from 
a flat Dalitz distribution.
For real data, there will almost certainly be some features present in the 
shape of the background; 
however, for our purposes, a flat background model should be sufficient. 
We take the expected background yields in each of the six final states to be 
equivalent. This should be a reasonable approximation provided the main source
of background events is from combinatorics (a plausible assumption for LHCb).
The ratio of background to signal, $B/S$, is defined here as the ratio of 
the number of expected background events to the number of expected signal
events for the final state with the smallest expected signal yield.
We generate background events according to this prescription for the ratios
$B/S = 1,2,10,50$ and 100, and fit the data using the same procedure as in the
previous section.

\begin{figure*}[!thb]
  \includegraphics[width=0.3\textwidth]{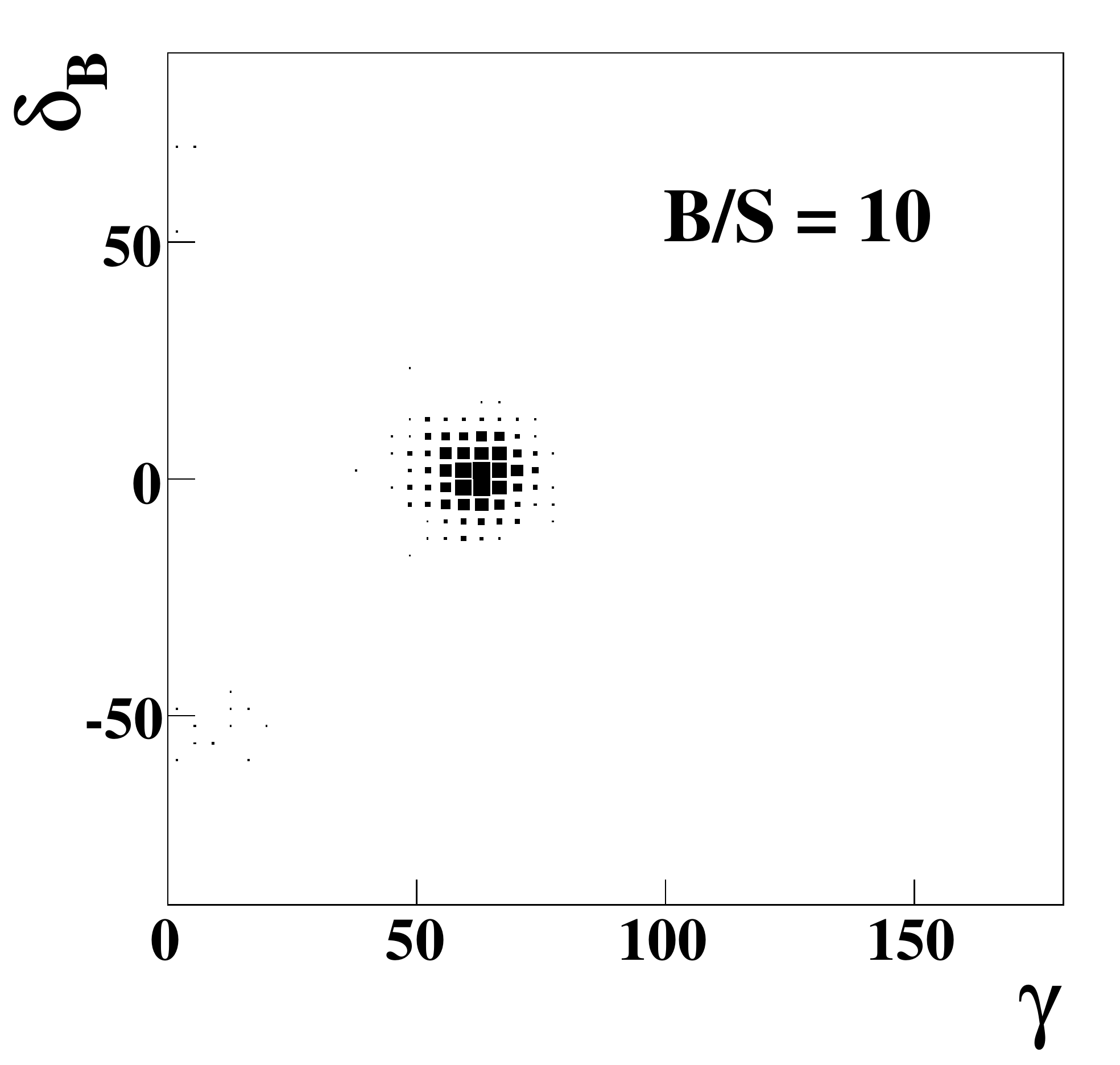}
  \includegraphics[width=0.3\textwidth]{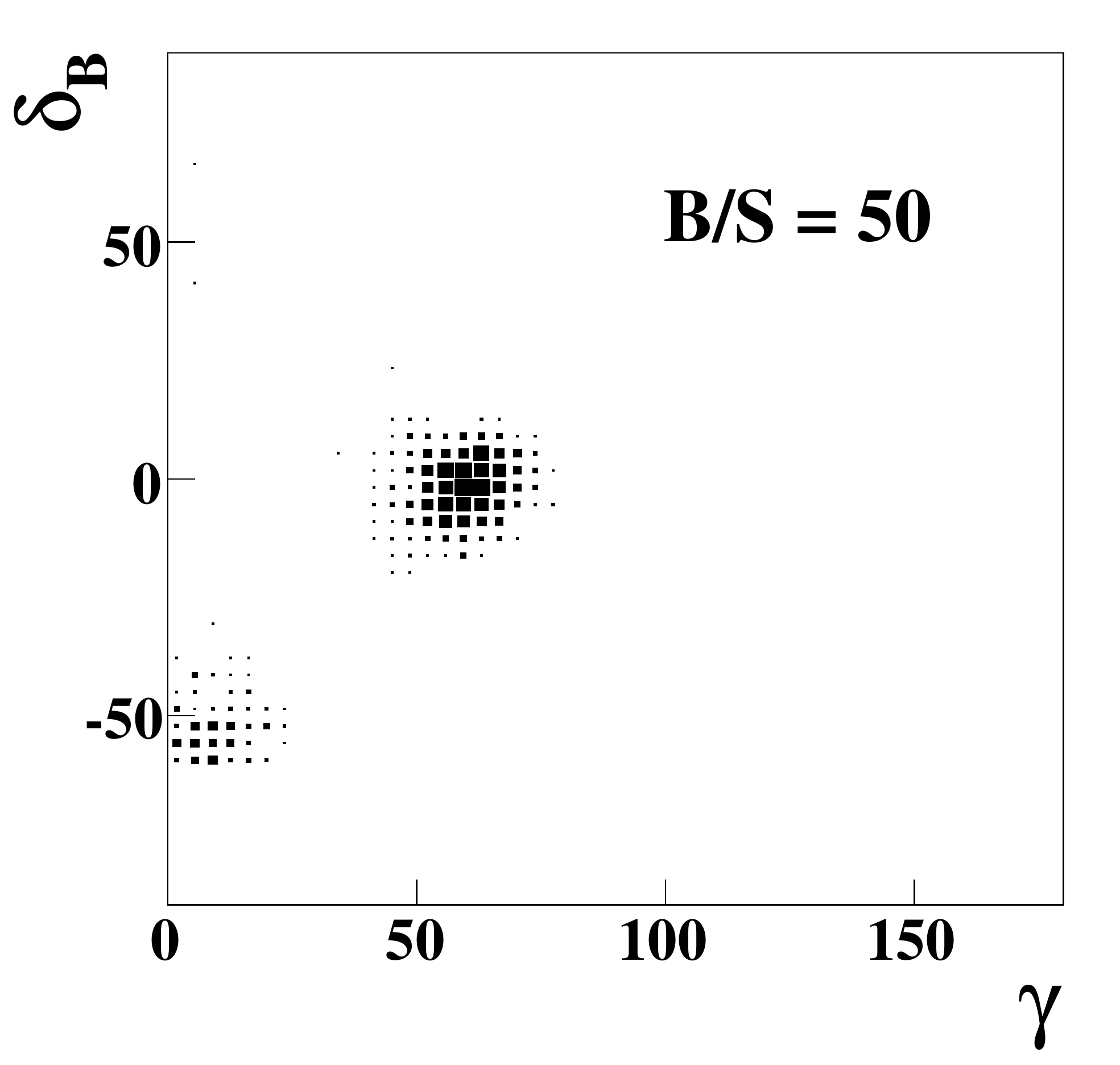}
  \includegraphics[width=0.3\textwidth]{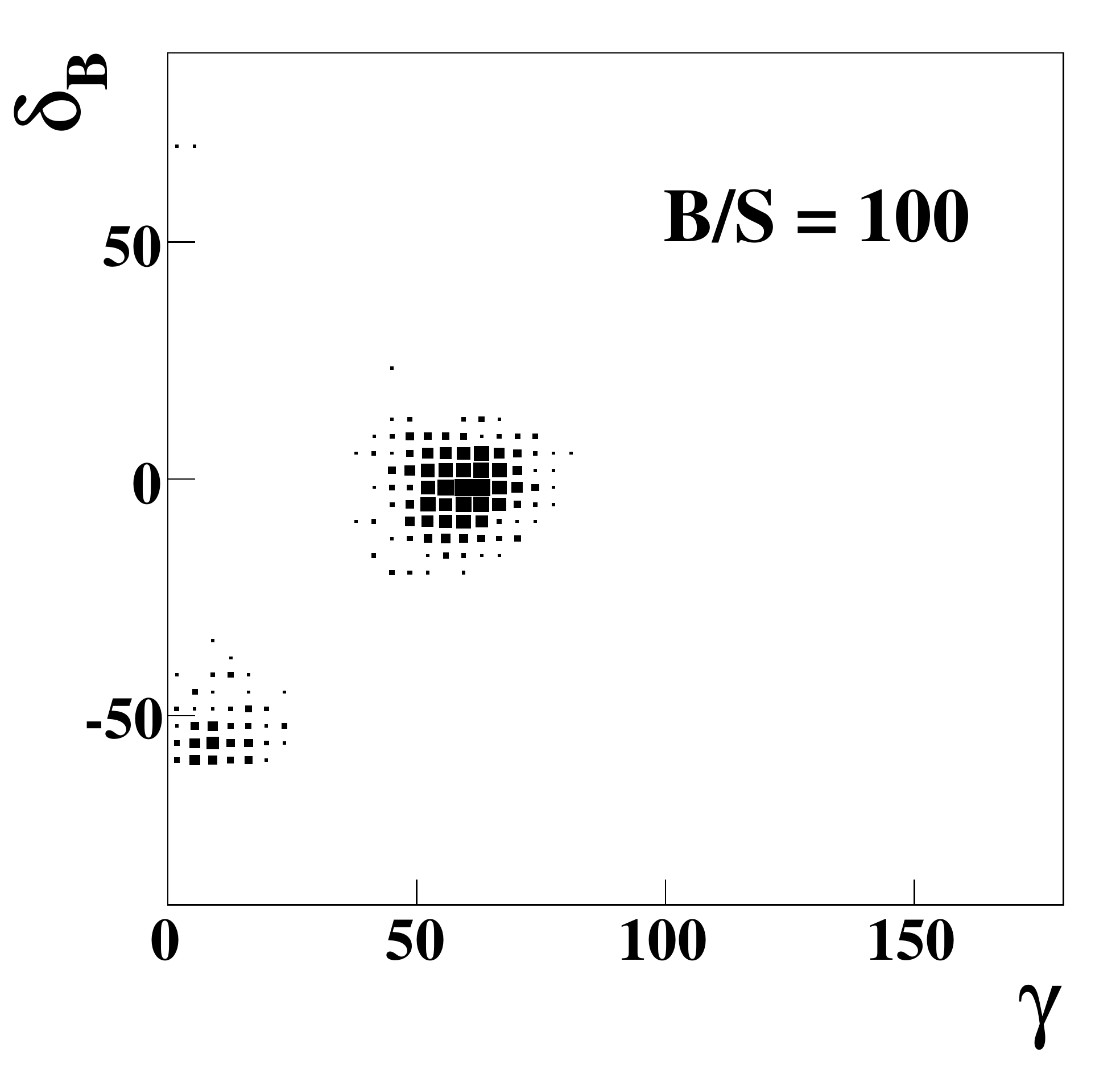}
  \caption{
    Distributions of fitted values of $\gamma$ in the Dalitz plot approach,
    for $B/S = 10,50,100$.
    These results are obtained with $\gamma = 60^\circ$,
    $\delta_B = 0^{\circ}$ and $r_B = 0.4$. 
  }
  \label{fig:bkgd}
\end{figure*}

For $B/S \gsim 10$, ambiguities in the solution begin to appear 
(see Fig.~\ref{fig:bkgd}),
just as they did using the quasi-two-body approach without background.
About 2\% of pseudo-experiments with $B/S=10$ find their best likelihood
in an ambiguous solution.  
This number steadily increases as the background increases, reaching about
25\% for $B/S = 100$.  
This can be understood since as the background increases, it becomes more
difficult to extract cleanly the amplitudes with smaller contributions.
The amplitude with the strongest contribution in our model is $DK^*(892)$.  
Thus, in the presence of a large background contamination, the amplitude 
analysis essentially reduces to the quasi-two-body approach.  
This results in the ambiguities in the solution found in the quasi-two-body 
approach (without background) appearing in the amplitude analysis if the 
background yields are large.

Our studies show that the background level should be kept to $B/S \lsim 20$
in order to take full advantage of the benefits of the amplitude analysis. 
This appears achievable, even in a hadronic environment.
For $B/S=20$ the ambiguities in the solution still appear at 
the few percent level; 
however, one can study the likelihood contours to determine whether the 
solution for any given experiment may have ambiguities.  
The results obtained from an experiment with such a background would need to 
take this into account.
Note also that these results are for a data sample roughly equivalent to one
nominal year's data taking at LHCb, and that effects due to ambiguities would
be expected to be ameliorated with larger data samples.

The results obtained for $r_B(DK^*(892))$, $\delta_B(DK^*(892))$ and $\gamma$
are given in Tab.~\ref{tab:res-bkgd}.  The ambiguities in the solution 
discussed above have been ignored in these results, {\em i.e.} the means and
widths for $B/S \geq 10$ are obtained by fitting the region around the true 
solution to a Gaussian lineshape.
The effects of the background on the resolution are minimal.  
The resolution of $\gamma$ for $B/S = 10$ is only $1^{\circ}$ worse than with
no background.  Even for  $B/S = 100$ the resolution is only about 60\% worse.
Thus, the most important impact of the presence of background events is 
their ability to lessen the power the amplitude analysis has to break the
ambiguities found in the quasi-two-body analysis.  

\begin{table}[!htb]
  \caption{
    Results for $r_B(DK^*(892))$, $\delta_B(DK^*(892))$ and $\gamma$
    obtained from the Dalitz-plot analysis, 
    with different levels of background. 
    These results are for $r_B = 0.4$, $\delta_B = 0^{\circ}$ and 
    $\gamma = 60^{\circ}$.
  }
  \label{tab:res-bkgd}
  \begin{tabular}{c|cc|cc|cc}
    \hline
    & \multicolumn{2}{c|}{$r_B$} & \multicolumn{2}{c|}{$\delta_B$ ($^{\circ}$)}
    & \multicolumn{2}{c}{$\gamma$ ($^{\circ}$)} \\
    & $\mu$ & $\sigma$ & $\mu$ & $\sigma$ & $\mu$ & $\sigma$ \\
    \hline
    $B/S = 0$   & 0.40 & 0.01 &  0.8 & 4.9 & 59.8 & 4.4 \\
    $B/S = 1$   & 0.40 & 0.01 &  0.1 & 5.1 & 59.8 & 4.9 \\
    $B/S = 2$   & 0.39 & 0.01 & -0.8 & 5.2 & 61.0 & 5.0 \\
    $B/S = 10$  & 0.40 & 0.01 &  1.2 & 5.2 & 62.1 & 5.4 \\
    $B/S = 50$  & 0.40 & 0.01 & -1.5 & 5.8 & 59.5 & 6.2 \\
    $B/S = 100$ & 0.39 & 0.01 & -1.9 & 6.0 & 58.8 & 7.2 \\
    \hline
  \end{tabular}
\end{table}

Another experimental effect which must be taken into account in a real
experimental analysis is the variation of reconstruction efficiency across the
Dalitz plot.  Our study has assumed that the efficiency is flat, but in
reality one would expect the probability to reconstruct successfully a decay
to be lower at the edges of phase space near the corners of the Dalitz plot.
If the shape of the efficiency function is known, this can be taken into
account in the analysis.  However, systematic effects arise since the
efficiency is typically measured using Monte Carlo simulations, which
inevitably will not give exactly the same behaviour as the data.

It is impossible to give a quantitative estimate of how large an effect this
may be.  Nonetheless, there is a strong qualitative reason to believe that it
will not be major problem.  The key point is that the our amplitude analysis
extracts $\gamma$ from the difference between $DK\pi$ Dalitz-plot
distributions with the $D$ meson reconstructed in different decay modes 
($D\to K\pi$ and $D \to CP$ eigenstates).
Data/MC differences can be expected to cancel in the ratio of Dalitz-plot
distributions, unless there is a momentum dependence in the ratio of
efficiencies to reconstruct the $D$ meson in the different final states.
Such an effect should be possible to study using control samples.

\section{Summary}
\label{sec:summary}

We have presented a feasibility study of an extension to the recently proposed
method to extract the Unitarity Triangle angle $\gamma$ from amplitude
analysis of $B \to DK\pi$ Dalitz plots.
The analysis includes the cases where the neutral $D$ meson is reconstructed
in $CP$-even eigenstates as well as in CKM-favoured and CKM-suppressed
hadronic decays. 
Compared to the previously proposed quasi-two-body analysis, 
the amplitude analysis provides 
(i) at least $50\%$ better sensitivity to $\gamma$, 
(ii) resolution of ambiguous solutions,
(iii) much reduced dependence of the sensitivity on the strong phase
$\delta_B$, and 
(iv) the possibility to determine the poorly known parameter $\delta_D$.
The analysis appears to be relatively robust against mismodelling of the
Dalitz plot, and performs well even when relatively large backgrounds are
present.  
We conclude that this method appears to be a highly attractive addition to the
family of methods that can be used to determine $\gamma$.

We are grateful to our colleagues from the LHCb experiment,
and would particularly like to thank Ignacio Bediaga, Alex Bondar,
Ulrik Egede,
Patrick Koppenburg, Anton Poluektov and Guy Wilkinson for discussions.
This work is supported by the
Science and Technology Facilities Council (United Kingdom).

\end{document}